\documentclass[12pt]{article} 
\usepackage{axodraw,cite} 
 
\def\ltsim{\lower3pt\hbox{$\, \buildrel < \over \sim \, $}} 
\def\gtsim{\lower3pt\hbox{$\, \buildrel > \over \sim \, $}} 
\def\be{\begin{equation}} 
\def\ee{\end{equation}} 
\def\ba{\begin{eqnarray}} 
\def\ea{\end{eqnarray}} 
\def\ga{\mathrel{\raise.3ex\hbox{$>$\kern-.75em\lower1ex\hbox{$\sim$}}}} 
\def\la{\mathrel{\raise.3ex\hbox{$<$\kern-.75em\lower1ex\hbox{$\sim$}}}}

\begin{document}

\baselineskip=16pt  
\begin{titlepage} 
\rightline{OUTP-01-42P} 
\rightline{hep-ph/0107307} 
\rightline{July 2001}   
\begin{center} 
 
\vspace{0.5cm} 
 
\large {\bf Multi-Localization in Multi-Brane Worlds} 
\vspace*{5mm} 
\normalsize 
 
{\bf Ian I. Kogan\footnote{i.kogan@physics.ox.ac.uk}, Stavros 
Mouslopoulos\footnote{s.mouslopoulos@physics.ox.ac.uk},  Antonios 
Papazoglou\footnote{a.papazoglou@physics.ox.ac.uk}}  
\\ and {\bf Graham 
G. Ross\footnote{g.ross@physics.ox.ac.uk}}

\smallskip  
\medskip  
{\it Theoretical Physics, Department of Physics, Oxford University}\\ 
{\it 1 Keble Road, Oxford, OX1 3NP,  UK} 
\smallskip

\vskip0.6in \end{center} 
  
\centerline{\large\bf Abstract} 
\smallskip 
\smallskip

We study bulk fields in various multi-brane models with localized 
gravity. In the case of spin $0$, $\frac{1}{2}$ and $\frac{3}{2}$  
fields, the non-trivial background geometry itself can induce 
(multi-) localization of the zero modes along the extra dimension. The addition of appropriate mass terms  
controls the strength or/and the position of this localization  and  
can induce (multi-) localization of spin $1$ fields as well. 
We show that multi-brane models may also give rise to anomalously 
light KK modes which are also localized. 
The possibility of  multi-localization in 
the context of supersymmetric brane world models in $AdS_{5}$ 
spacetime is also discussed.

\vspace*{2mm}  

\end{titlepage} 
 
\section{Introduction} 
 
Brane universe models \cite{Akama:1982jy,Rubakov:1983bb,Visser:1985qm,Squires:1986aq} in more than four 
dimensions have been extensively studied over the last three years because 
they provide novel ideas for resolutions of long standing problems of 
particle physics such as the Planck hierarchy \cite{Arkani-Hamed:1998rs,Antoniadis:1998ig,Arkani-Hamed:1999nn,Gogberashvili:1998vx,Randall:1999ee}%
. Moreover, mechanisms to localize gravity on a brane \cite{Gogberashvili:1998vx,Randall:1999ee} have led to the realization that if 
extra dimensions exist, they need not be compact \cite{Randall:1999vf}. In 
the simplest formulation of these models no bulk matter states are assumed 
to exist and thus only gravity propagates in the extra dimensions. 
Nevertheless ``bulk'' (\textit{i.e.} transverse to 3-brane space dimensions) 
physics turns out to be very interesting giving alternative explanations to 
other puzzles of particle physics. For example, by assuming the existence of 
a Standard Model (SM) neutral spin $\frac{1}{2}$ fermion in the bulk one can 
explain the smallness of the neutrino masses without invoking the seesaw 
mechanism \cite{Dienes:1999sb,Arkani-Hamed:1998vp,Dvali:1999cn,Mohapatra:1999zd,Barbieri:2000mg,Lukas:2000wn,Lukas:2000rg,Cosme:2000ib,Grossman:2000ra,Mouslopoulos:2001uc}. However, it is not necessary to confine the SM fields to the brane. 
Assuming that the SM fields can propagate in the bulk interesting new 
possibilities arise. For example one can attempt to explain the pattern of 
the SM fermion mass hierarchy by localizing the SM fermions at different 
places in the bulk \cite{Arkani-Hamed:2000dc,Mirabelli:2000ks,Dvali:2000ha,delAguila:2000kb}. These 
considerations give the motivation for considering the phenomenology 
associated with spin $0$, $\frac{1}{2}$ and $1$ fields propagating in extra 
dimension(s). Since our discussions will be limited to models with localized 
gravity and in particular to Randall-Sundrum (RS) type constructions, we 
will be interested in the phenomenology of fields that live in a slice of $%
AdS_{5}$ spacetime. If one also wants to explore the supersymmetric version 
of the above models, it is also necessary to study the phenomenology of spin  
$\frac{3}{2}$ field on the same background geometry. 
 
It has been shown that in the context of RS type models  the graviton  is 
localized on positive tension branes and suppressed on the negative ones  
\cite{Randall:1999ee}. In Ref.\cite{Goldberger:1999wh} it was also shown 
that the $AdS_{5}$ background geometry of these models can localize the zero 
mode of a massless scalar field on positive tension branes (see also \cite{Mintchev:2001mf}). Moreover, the same background localizes 
the spin $\frac{1}{2}$ fermions on negative tension branes \cite{Grossman:2000ra,Mouslopoulos:2001uc,Bajc:2000mh,Chang:2000nh,Gherghetta:2000qt,Randjbar-Daemi:2000cr,Oda:2000kh,Kehagias:2000au,Oda:2000dd,Gherghetta:2000kr}. The same localization behaviour holds for spin $\frac{3}{2}$ fermions \cite{Bajc:2000mh,Oda:2000kh,Oda:2000wa}. However, the $AdS_{5}$ background 
geometry cannot localize massless Abelian gauge fields \cite{Pomarol:2000ad,Davoudiasl:2000tf,Davoudiasl:2001wi} if one tries to get them from a higher
dimensional vector field. One can, however, get localized massless Abelian
fields from higher dimensional antisymmetric forms as it was done in
\cite{Duff:2001jk}.

The addition of mass terms modifies the localization properties. For 
example, it was shown in Ref.\cite{Rubakov:1983bb,Jackiw:1976fn,Grossman:2000ra,Kehagias:2000au,Mouslopoulos:2001uc} 
that the addition of an appropriate mass term in the action of a spin $\frac{%
1}{2}$ field can result in the localization of the zero mode that resembles 
that of the graviton (the magnitude of the mass term in this case controls 
the extent of localization). In the present paper we show that the same can 
occur in the case of spin $0$, $\frac{3}{2}$ fields with appropriate mass 
terms. Furthermore we show that by adding a mass term of a particular form 
in the action of a massless Abelian gauge field we can achieve the desired 
localization of the massless zero mode which can be made to resemble that of 
the graviton. The mass term in this case must necessarily consist of a five 
dimensional bulk mass part and a boundary part. 
 
From the above it is clear that particles of all spins, with appropriate 
bulk mass terms, can exhibit zero mode localization on positive tension 
branes, just as for the graviton. In the context of multi-brane models with 
localized gravity the above implies a further interesting possibility: the 
phenomenon of multi-localization in models that contain at least two 
positive tension branes. Multi-localization, as we will see, is closely 
related to the appearance of light, localized strongly coupled
Kaluza-Klein (KK) states 
(their coupling to matter can be even larger than the coupling of the zero 
mode). Thus the mass spectrum of multi-localized fields is distinct from the 
mass spectrum of singly localized fields, resulting in the possibility of 
new phenomenological signals. Anomalously light states may also arise in 
theories without multi-localization (\textit{i.e.} even in configurations 
with one positive brane) in models with twisted boundary conditions. 
 
The appearance of light KK states can be of particular phenomenological 
interest. For example in the case of the graviton, multi-localization and 
thus the appearance of light KK graviton excitations, gives rise to the 
exciting possibility of Bi-gravity \cite{Kogan:2000wc,Mouslopoulos:2000er,Kogan:2001vb,Kogan:2001yr} (Multi-gravity  
\cite{Kogan:2001yr,Gregory:2000jc,Kogan:2000cv,Kogan:2000xc}) where part (or even all) of 
gravitational interactions can come from massive spin $2$ particle(s) (KK 
state(s)) \cite{Kogan:2000wc,Mouslopoulos:2000er,Kogan:2001yr,Kogan:2001vb,Gregory:2000jc,Kogan:2000cv,Kogan:2000xc,Karch:2001ct,Miemiec:2000eq,Schwartz:2001ip,Karch:2001cw}. In this case the large mass gap between the anomalously light KK state(s) 
and the rest of the tower is critical in order to avoid modifications of 
Newton's law at intermediate distances. 
 
Anomalously light spin $\frac{1}{2}$ KK states can also arise when a bulk 
fermion is multi-localized \cite{Mouslopoulos:2001uc}. The non-trivial 
structure of the KK spectrum in this case, with the characteristic mass gap 
between the light state(s) and the rest of the tower, can be used for 
example to construct models with a small number of active or sterile 
neutrinos involved in the oscillation (the rest will decouple since they 
will be heavy). 
 
 
The organization of the paper is as follows: In the next Section we review 
the general framework and discuss the general idea of multi-localization in 
the context of models with localized gravity. In Section 3 we study the 
multi-localization properties of a bulk scalar field. In Section 4 we review 
the situation of a bulk fermion field. In Section 5 we study in detail the 
possibilities of localization and multi-localization of an Abelian gauge 
field. In Section 6 present the possibility of multi-localization in the 
case of a gravitino and finally in Section 7 we review the same phenomena 
for the case of graviton \cite{Kogan:2000wc,Mouslopoulos:2000er}. In section 
8 we discuss how multi-localization is realized in the context of 
supersymmetric versions of the previous models. The overall implications and 
conclusions are presented in section 9. 
 
\section{General Framework - The idea of Multi-Localization} 
 
The original formulation of multi-localization of gravity was obtained in 
five dimensions for the case that there is more than one positive tension 
brane. If the warp factor has a ``bounce'', in the sense that it has a 
minimum (or minima) between the positive tension branes, the massless 
graviton appears as a bound state of the attractive potentials associated 
with the positive tension branes with its wave function peaked around them. 
Moreover in this case there are graviton excitations corresponding to 
additional bound states with wave functions also peaked around the positive 
tension branes. They are anomalously light compared to the usual Kaluza 
Klein tower of graviton excitations. The reason for this is that the 
magnitude of their wavefunction closely approximates that of the massless 
mode, differing significantly only near the position of the bounce where the 
wave function is exponentially small. The mass they obtain comes from this 
region and as a result is exponentially suppressed relative to the usual KK 
excitations. 
 
The first models of this type involved negative tension branes sandwiched 
between the positive tension branes \cite{Kogan:2000wc}. That this is 
necessary in the case of flat branes with vanishing cosmological constant in 
four dimensions is easy to see because, for a single flat brane, the minimum 
of the warp factor is at infinity and thus any construction with another 
positive brane at a finite distance will have a discontinuity in the 
derivative of the warp factor at the point of matching of the solutions and 
thus at that point a negative tension brane will emerge. 
 
Subsequently it was pointed out that a free negative tension brane(s) 
violate the weaker energy condition and lead to a ghost radion field(s). A 
way out of this problem was suggested by some of us through the use of
$AdS_{4}$ branes \cite{Kogan:2001vb} (see \cite{Gorsky:2000rz} for a
different possibility involving an external four-form field). In this case the minimum of a single 
brane is at finite distance and thus one can match the solution for two 
positive branes without introducing a negative tension brane. However a 
drawback of this approach is that the four dimensional cosmological constant 
is negative in conflict with the current indications for a positive 
cosmological constant. 
 
Recently we have shown how it is possible to obtain the ``bounce'' and the 
related multilocalisation for zero cosmological constant in four dimensions 
without the need for negative tension branes \cite{Kogan:2001yr}. We have 
shown this is possible even in cylindrically symmetric models if one allows 
for non-homogeneous brane tensions and/or bulk cosmological constant. In 
this case the structure of the effective four dimensional theory is very 
similar to the five dimensional case for the modes which do not depend on 
the new angular co-ordinate. In particular one finds a massless graviton and 
anomalously light massive modes with wavefunctions peaked around the 
positive tension branes. 
 
In this paper we are interested in whether spin $0,\frac{1}{2},1$, and $%
\frac{3}{2}$ fields can similarly show the phenomena of multi-localisation 
we found for the graviton in suitable curved backgrounds. We will show that 
the curved background can also induce localisation for the case of spin $0,%
\frac{1}{2}$ and $\frac{3}{2}$ fields but not for a massless vector field. 
However, even in flat spacetime in higher dimensions, it is also possible to 
induce localisation by introducing mass of a very specific form for these 
fields. We will show that this effect can localise all fields with spin $%
\leq \frac{3}{2},$ including the graviton. 
 
For simplicity we will work mainly with the five dimensional 
compactification. In the context of the discussion of the multilocalisation 
of fields of spin $\leq \frac{3}{2},$ what is important is the nature of the 
curved background as determined by the warp factor. Thus the general 
features are applicable to all models of multilocalisation with 
the same warp factor profile. However the interpretation of the mass terms 
needed to achieve multilocalisation differs. In the case of models with 
negative tension branes the masses correspond to a combination of a constant 
bulk mass together with a brane term corresponding to the coupling to 
boundary sources. In the case of models without negative tension branes the 
mass must have a non-trivial profile in the bulk. In certain cases this 
profile may be guaranteed by supersymmetry. 
 
In the five dimensional models considered here, the fifth dimension $y$ is 
compactified on an orbifold, $S^{1}/Z_{2}$ of radius $R$, with $-L \le y \le 
L $. The five dimensional spacetime is a slice of $AdS_{5}$ which is 
described by\footnote{%
We will assume that the background metric is not modified by the presence of 
the bulk fields, that is, we will neglect the back-reaction on the metric 
from their presence.}:  
\begin{equation} 
ds^{2}=e^{-2 \sigma(y)} \eta_{\mu \nu}dx^{\mu}dx^{\nu}+dy^2 
\end{equation} 
where the warp factor $\sigma(y)$ depends on the details of the model 
considered. 
 
Since we are interested in the phenomenology of fields propagating in the 
above slice of $AdS_{5}$ our goal is to determine the mass spectrum and 
their coupling to matter. Starting from a five dimensional Lagrangian, in 
order to give a four dimensional interpretation to the five dimensional 
fields, one has to implement the dimensional reduction. This procedure 
includes the representation of the five dimensional fields $\Phi (x,y)$ in 
terms of the KK tower of states:  
\begin{equation} 
\Phi (x,y)=\sum_{n=0}^{\infty }\Phi ^{(n)}(x)f^{(n)}(y) 
\end{equation} 
where $f^{(n)}(y)$ is a complete orthonormal basis spanning the compact 
dimension. The idea behind this KK decomposition is to find an equivalent 4D 
description of the five dimensional physics associated with the field of 
interest, through an infinite number of KK states with mass spectrum and 
couplings that encode all the information about the five dimensions. The 
function $f^{(n)}(y)$ describes the localization of the wavefuntion of the 
n-th KK mode in the extra dimension. It can be shown that $f^{(n)}(y)$ obeys 
a second order differential equation which, after a convenient change of 
variables and/or a redefinition\footnote{The form of the redefinitions depend on the spin of the field.} of the 
wavefunction, reduces to an ordinary Schr\"{o}dinger equation:  
\begin{equation} 
\left\{ -\frac{1}{2}\partial _{z}^{2}+V(z)\right\} \hat{f}^{(n)}(z)=\frac{%
m_{n}^{2}}{2}\hat{f}^{(n)}(z) 
\end{equation} 
The mass spectrum and the wavefunctions (and thus the couplings) are 
determined by solving the above differential equation. Obviously all the 
information about the five dimensional physics is contained in the form of 
the potential $V(z)$. For example in the case of the graviton the positive 
tension branes correspond to attractive $\delta $-function potential wells 
whereas negative tension branes to $\delta $-function barriers. The form of 
the potential between the branes is determined by the $AdS_{5}$ background. 
 
\subsection{Multi-Localization and light KK states} 
 
 
\begin{figure}[t] 
\begin{center} 
\SetScale{0.7}  
\begin{picture}(200,150)(0,50) 
\LongArrow(140,0)(140,280) 
\LongArrow(-80,120)(370,120) 
 
\SetWidth{1} 
 
\SetColor{Green} 
\Line(140,20)(148,250) 
\Line(140,20)(132,250) 
\SetColor{Black} 
 
\SetColor{Red} 
\Line(240,270)(245,130) 
\Line(240,270)(235,130) 
\Line(40,270)(45,130) 
\Line(40,270)(35,130) 
\SetColor{Black} 
 
\SetColor{Green} 
\Line(-15,140)(-20,0) 
\Line(-25,140)(-20,0) 
\Line(305,140)(300,0) 
\Line(295,140)(300,0) 
\SetColor{Black} 
 
\SetColor{Black} 
 
 
\Text(100,200)[l]{$V(z)$} 
\Text(260,70)[rb]{$z$} 
 

\SetColor{Black} 
\Curve{(148,250)(153,249)(170,190)(200,138)(220,131)(235,130)} 
\Curve{(245,130)(260,131)(295,140)}

\Curve{(45,130)(60,131)(80,138)(110,190)(127,249)(132,250)} 
\Curve{(-15,140)(20,131)(35,130)}

\end{picture} 
\end{center} 
\par 
\vspace*{8mm} 
\caption{The scenario of multi-localization is realized in configurations 
where the corresponding form of potential has potential wells that can 
support bound states. Such a potential is the one that corresponds to the $%
^{\prime\prime}+-+^{\prime\prime}$ model. Positive branes are $\protect\delta 
$-function wells and negative are $\protect\delta$-function barriers.} 
\end{figure}
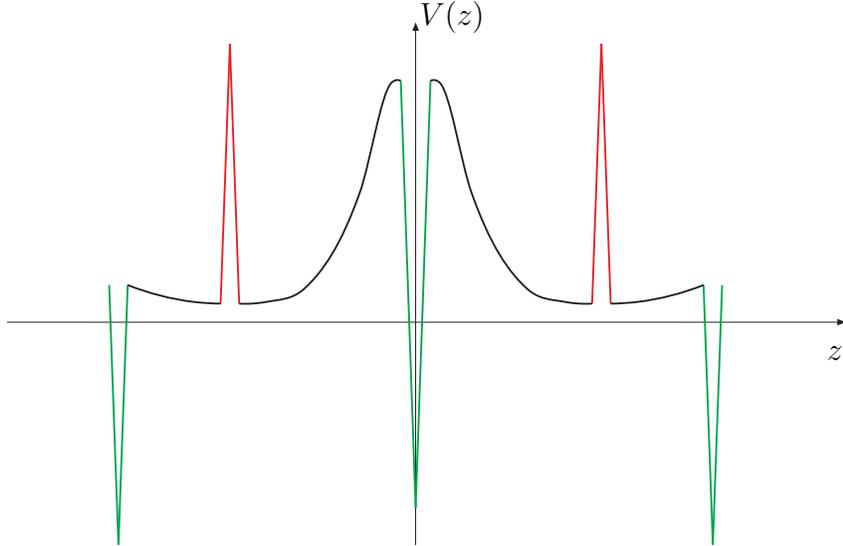 
 
Multi-localization emerges when one considers configuration of branes such 
that the corresponding potential $V(z)$ has at least two ($\delta $-function)%
\footnote{In the infinitely thin brane limit that we consider, the wells associated 
with positive branes are $\delta $-functions.} potential wells, each of which 
can support a bound state (see Fig.(1) for the $^{\prime \prime }+-+^{\prime 
\prime }$ case). If we consider the above potential wells separated by an 
infinite distance, then the zero modes are degenerate \textit{i.e.} and 
massless. However, if the distance between them is finite, due to quantum 
mechanical tunneling the degeneracy is removed and an exponentially small 
mass splitting appears between the states. The rest of levels, which are not 
bound states, exhibit the usual KK spectrum with mass difference 
exponentially larger than the one of the ``bound states'' (see Fig.(2)). The 
above becomes clearer if one examines the form of the wavefunctions. In the 
finite distance configuration the wavefunction of the zero mode is the 
symmetric combination ($\hat{f}_{0}=\frac{\hat{f}_{0}^{1}+\hat{f}_{0}^{2}}{%
\sqrt{2}}$) of the wavefunctions of the zero modes of the two wells whereas 
the wavefunction of the first KK state is the antisymmetric combination ($%
\hat{f}_{0}=\frac{\hat{f}_{0}^{1}-\hat{f}_{0}^{2}}{\sqrt{2}}$). Such an 
example is shown in Fig.(3) where are shown the wavefunctions of the 
graviton in the context of $^{\prime \prime }+-+^{\prime \prime }$ model 
with two positive tension branes at the fixed point boundaries and a 
negative tension brane at the mid-point. From it we see that the absolute 
value of these wavefunctions are nearly equal throughout the extra 
dimension, with exception of the central region where the antisymmetric 
wavefunction passes through zero, while the symmetric wavefunction has 
suppressed but non-zero value. The fact that the wavefunctions are 
exponentially small in this central region results in the exponentially 
small mass difference between these states. 
 
The phenomenon of multi-localization is of particular interest since, 
starting from a problem with only one mass scale (the inverse radius of 
compactification), we are able to create a second scale exponentially 
smaller. Obviously the generation of this hierarchy is due to the tunneling 
effects in our ``quantum mechanical'' problem. 
 
 
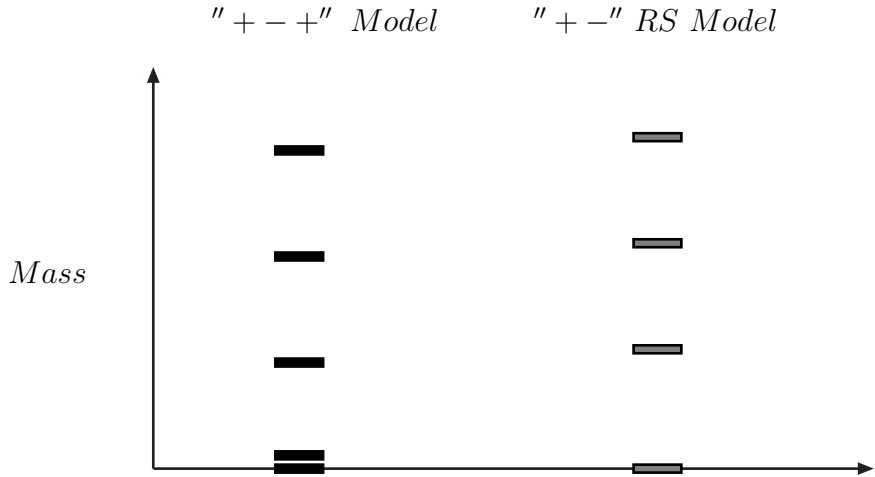
\begin{figure}[t] 
\begin{center} 
\begin{picture}(250,200)(0,70) 
 
\SetWidth{1} 
\LongArrow(-20,100)(-20,250) 
 
\SetWidth{1} 
\LongArrow(-20,100)(250,100) 

\Text(-60,175)[c]{$Mass$} 
\Text(45,270)[c]{$''+-+''~Model$} 
\Text(170,270)[c]{$''+-''~RS~Model$} 
 
 
 
\GBoxc(35,220)(18,3){0} 
\GBoxc(35,180)(18,3){0} 
\GBoxc(35,140)(18,3){0} 
 
\GBoxc(35,105)(18,3){0} 
 
\GBoxc(35,100)(18,3){0} 
 
 
 

\GBoxc(170,225)(18,3){0.5} 
\GBoxc(170,185)(18,3){0.5} 
\GBoxc(170,145)(18,3){0.5} 
 
 
\GBoxc(170,100)(18,3){0.5} 
 
 
\end{picture} 
\end{center} 
\caption{Comparison of the gravitational spectrum of the $^{\prime \prime 
}+-+^{\prime \prime }$ model with the $^{\prime \prime \prime }+-^{\prime }$ 
Randall-Sundrum model.} 
\end{figure} 
 

\begin{figure}
\begin{center}

\SetScale{0.5}
\begin{picture}(180,100)(0,50)

\SetWidth{2}
\Line(10,50)(10,250)
\Line(290,50)(290,250)

\SetWidth{0.5}
\Line(150,50)(150,250)
\Line(10,150)(290,150)



{\SetColor{Green}
\Curve{(10,240)(50,192)(65,181)(80,173)(95,167)(110,162)(130,157)(150,155)}
\Curve{(150,155)(170,157)(190,162)(205,167)(220,173)(235,181)(250,192)(290,240)}
}


{\SetColor{Red}
\DashCurve{(10,240)(50,192)(65,181)(80,173)(95,167)(110,161)(130,155)(150,150)}{3}
\DashCurve{(150,150)(170,145)(190,139)(205,133)(220,127)(235,119)(250,108)(290,60)}{3}
}


{\SetColor{Brown}
\DashCurve{(10,145)(15,149)(20,150)(40,153)(65,158)(140,209)(150,210)}{1}
\DashCurve{(150,210)(160,209)(235,158)(260,153)(280,150)(285,149)(290,145)}{1}
}
\end{picture}

\end{center}

\vspace*{3mm} 

\caption{The wavefunctions of the three first  modes in  the $''+-+''$ 
model. The
zero mode (solid line), first (dashed line )  and second (dotted line)
KK states. Note that the absolute value of the wavefunctions of the
zero mode and the first KK state almost coincide except for the
central region of the configuration where they are both suppressed.
}


\vspace*{3mm}


\begin{center}

\SetScale{0.5}

\begin{picture}(180,100)(0,50)

\SetWidth{2}
\Line(-30,50)(-30,250)
\Line(330,50)(330,250)

\SetWidth{2.5}
\LongArrow(-70,150)(-110,150)
\LongArrow(370,150)(410,150)

\SetWidth{0.5}
\Line(150,50)(150,250)
\Line(-30,150)(330,150)



{\SetColor{Green}
\Curve{(-30,240)(10,192)(25,181)(40,173)(55,167)(70,162)(90,157)(110,155)(130,155)(150,155)(170,155)(190,155)}
\Curve{(190,155)(210,157)(230,162)(245,167)(260,173)(275,181)(290,192)(330,240)}
}


{\SetColor{Red}
\DashCurve{(-30,240)(10,192)(25,181)(40,173)(55,167)(70,161)(90,155)(110,155)(130,152)(150,150)}{3}
\DashCurve{(150,150)(170,147)(190,145)(210,145)(230,139)(245,133)(260,127)(275,119)(290,108)(330,60)}{3}
}


{\SetColor{Brown}
\DashCurve{(-30,145)(-10,145)(0,147)(15,149)(20,150)(40,153)(65,158)(140,200)(150,201)}{1}
\DashCurve{(150,201)(160,200)(235,158)(260,153)(280,150)(285,149)(290,147)(300,147)(310,145)(330,145)}{1}
}
\end{picture}

\end{center}

\vspace*{3mm} 

\caption{The effect of stretching the configuration by moving the two wells further
apart. Note that the wavefunction of the zero mode and the first KK
state remain localized but the remaining of the modes,  not being
bound states, will stretch along the extra dimension.}



\begin{center}

\SetScale{0.5}

\begin{picture}(180,100)(0,50)

\SetWidth{2}
\Line(-120,50)(-120,250)
\Line(420,50)(420,250)

\SetWidth{0.5}
\Line(-120,150)(130,150)
\Line(170,150)(420,150)

\SetWidth{2.5}
\LongArrow(-140,150)(-180,150)
\LongArrow(440,150)(480,150)

\SetWidth{0.5}

\Text(76,75)[c]{$\dots$}


{\SetColor{Green}
\Curve{(-120,240)(-80,192)(-65,181)(-50,173)(-35,167)(-20,162)(0,157)(40,155)(60,155)(80,155)}
\Curve{(200,155)(250,155)(300,157)(320,162)(335,167)(350,173)(365,181)(380,192)(420,240)}
}


{\SetColor{Red}
\DashCurve{(-120,240)(-80,192)(-65,181)(-50,173)(-35,167)(-20,162)(0,157)(40,155)(60,155)(80,155)}{3}
\DashCurve{(200,145)(250,145)(300,143)(320,138)(335,133)(350,127)(365,119)(380,108)(420,60)}{3}
}


{\SetColor{Brown}
\DashCurve{(-120,145)(-80,155)(-65,155)(-50,155)(-35,155)(-20,155)(0,155)(40,155)(60,155)(80,155)}{1}
\DashCurve{(200,155)(250,155)(300,155)(320,155)(335,155)(350,155)(365,155)(380,155)(420,145)}{1}
}
\end{picture}

\end{center}

\vspace*{3mm} 

\caption{ The case when the distance between the branes become infinite. The zero
mode (which still exists if the compactification volume is finite) and
the first KK state become degenerate. The wavefunction of the second KK
state spreads along the extra dimension.}

\end{figure}

 
\subsection{Locality - light KK states and separability} 
 
Summarizing, if in a single brane configuration (with infinite extra 
dimension) the zero mode of a field is localized on the brane, then in a 
multi-brane world scenario it will be multi-localized and as a consequence 
light KK states will appear in its spectrum. The latter is assured from the 
following locality argument: In the infinite separation limit of the 
multi-brane configuration physics on each brane should depend on local 
quantities and not on physics at infinity. The latter assures the smoothness 
of the limit of infinite separation of branes in the sense that at the end 
of the process the configuration will consists of identical and independent 
single brane configurations. As a result the above locality argument also 
ensures the appearance of light states which at the above limit will become 
the zero modes of the one brane configurations. 
 
\section{Multi-Localization of spin 0 field} 
 
\subsection{$^{\prime\prime}+-+^{\prime\prime}$ Model} 
 
Let us now explore whether multi-localization of spin $0$ fields can be 
realized in the context of the RS type of models. We start our discussion 
from the simplest case of a real scalar field propagating in a five 
dimensional curved background described by the metric of eq.(1) where the 
function $\sigma (y)$ is the one that corresponds to the $^{\prime \prime 
}+-+^{\prime \prime }$ configuration. The action for a massive bulk scalar 
field in this case is:  
\begin{equation} 
S=\frac{1}{2}\int d^{4}x\int dy\sqrt{G}\left( G^{AB}\partial _{A}\Phi 
\partial _{B}\Phi +m_{\Phi }^{2}\Phi ^{2}\right)  
\end{equation} 
where $G=det(G_{AB})=e^{-8\sigma (y)}$. Under the $Z_{2}$ symmetry $m_{\Phi 
}^{2}$ should be even. We take $m_{\Phi }^{2}$ of the form:  
\begin{equation} 
m_{\Phi }^{2}=C+\Sigma _{i}D_{i}\delta (y-y_{i}) 
\end{equation} 
where $D_{i}=\pm 1$ for a positive or negative brane respectively. The first 
term corresponds to a constant five dimensional bulk mass and the second to 
the coupling of the scalar field to boundary sources \footnote{%
Here we have assumed that the magnitude of the boundary mass term 
contribution is the same for all branes. This is needed in order to have a 
zero mode.}. The mass can be rewritten in the form  
\begin{equation} 
m_{\Phi }^{2}=\alpha (\sigma ^{\prime }(y))^{2}+\beta \sigma ^{\prime \prime 
}(y) 
\end{equation} 
Taking in account the form of the vacuum, the above action can be written as  
\begin{equation} 
S=\frac{1}{2}\int d^{4}x\int dy\left( e^{-2\sigma (y)}\eta ^{\mu \nu 
}\partial _{\mu }\Phi \partial _{\nu }\Phi -\Phi \partial _{5}(e^{-4\sigma 
(y)}\partial _{5}\Phi )+m_{\Phi }^{2}e^{-4\sigma (y)}\Phi ^{2}\right)  
\end{equation} 
In order to give a four dimensional interpretation to this action we go 
through the dimensional reduction procedure. Thus we decompose the five 
dimensional field into KK modes  
\begin{equation} 
\Phi (x,y)=\sum_{n}\phi _{n}(x)f_{n}(y) 
\end{equation} 
Using this decomposition, the above action can be brought in the form  
\begin{equation} 
S=\frac{1}{2}\sum_{n}\int d^{4}x\{\eta ^{\mu \nu }\partial _{\mu }\phi 
_{n}(x)\partial _{\nu }\phi _{n}(x)+m_{n}^{2}\phi _{n}^{2}(x)\} 
\end{equation} 
provided the KK wavefunctions obey the following second order differential 
equation  
\begin{equation} 
-\frac{d}{dy}\left( e^{-4\sigma (y)}\frac{df_{n}(y)}{dy}\right) +m_{\Phi 
}^{2}e^{-4\sigma (y)}f_{n}(y)=m_{n}^{2}e^{-2\sigma (y)}f_{n}(y) 
\end{equation} 
with the following orthogonality relations (taking in account the $Z_{2}$ 
symmetry):  
\begin{equation} 
\int_{-L}^{L}dye^{-2\sigma (y)}{f}_{m}^{\ast }(y)f_{n}(y)=\delta _{mn} 
\end{equation} 
where we assume that the length of the orbifold is $2L$. 
 
The linear second order differential equation can always be brought to a 
Schr\"ondiger form by a redefinition of the wavefunction and by a convenient 
coordinate transformation from y to z coordinates related by : $\frac{dz}{dy}%
=e^{\sigma(y)}$. The coordinate transformation is chosen to eliminate the 
terms involving first derivatives. Thus we end up with the differential 
equation of the form:  
\begin{equation} 
\left\{- \frac{1}{2}\partial_z^2+V(z)\right\}\hat{f}_{n}(z)=\frac{m_n^2}{2}%
\hat{f} _{n}(z) 
\end{equation} 
where the potential is given by  
\begin{equation} 
\hspace*{0.5cm} V(z)=\frac{\frac{15}{4}(\sigma^{\prime}(y))^{2}+m_{\Phi}^{2}%
}{2[g(z)]^2}- \frac{\frac{3}{2}}{2[g(z)]^2}\sigma^{\prime\prime}(y) 
\end{equation} 
where $g(z)\equiv e^{\sigma(y)}$ and we have made a redefinition of the 
wavefunction:  
\begin{equation} 
\hat{f}_{n}(z)=e^{-\frac{3}{2} \sigma(y)} f_{n}(y) 
\end{equation} 
 
Note that for $m_{\phi }=0$ the above Schr\"{o}endiger equation is identical 
to that of the graviton. This implies that the mass spectrum of a massless 
scalar field of even parity is identical to the graviton's and thus supports 
an ultralight KK state(s). Addition of a bulk mass term ($\beta =0$), 
results in the disappearance of the zero mode from the spectrum (the 
ultralight state also is lost). Nevertheless, by considering a mass term of 
the more general form (with $\alpha \neq 0$ and $\beta \neq 0$ ), which has 
the characteristic that it changes both terms of the potential of eq.(15), 
we can not only recover the zero mode but in addition have ultralight KK 
state(s). In this case the corresponding potential will be  
\begin{equation} 
\hspace*{0.5cm}V(z)=\frac{\left( \frac{15}{4}+\alpha \right) (\sigma 
^{\prime }(y))^{2}}{2[g(z)]^{2}}-\frac{\left( \frac{3}{2}-\beta \right) }{%
2[g(z)]^{2}}\sigma ^{\prime \prime }(y) 
\end{equation} 
This is of the general form given in Appendix A. A massless mode exists if $%
\alpha =\beta ^{2}-4\beta $ in which case the wavefunction is ${\hat{f}}%
(z)\propto e^{(\beta -3/2)\sigma (y)}$. From equation (11) we see that $%
f(y)e^{-\sigma (y)}\propto e^{(\beta -1)\sigma (y)}$ is the appropriately 
normalised wavefunction in the interval $[-L,L]$. This is localised on the 
positive tension brane for $\beta >1$ and on the negative tension brane for $%
\beta <1$. When the condition for the zero mode is satisfied we find the 
mass of the first ultralight KK state to be given by.  
\begin{equation} 
m_{1}\approx \sqrt{4{\nu }^{2}-1}~kw~e^{-(\nu +\frac{1}{2})x} 
\end{equation} 
where $\nu =\frac{3}{2}-\beta $ and for the rest of the KK tower  
\begin{equation} 
m_{n+1}\approx \xi _{n}~kw~e^{-x}~~~~~~n=1,2,3,\ldots  
\end{equation} 
where $\xi _{2i+1}$ is the $(i+1)$-th root of $J_{\nu -\frac{1}{2}}(x)$ ($%
i=0,1,2,\ldots $) and $\xi _{2i}$ is the $i$-th root of $J_{\nu +\frac{1}{2}%
}(x)$ ($i=1,2,3,\ldots $). The above approximations become better away from 
the $\nu =\frac{1}{2}$ , $x=0$ and for higher KK levels, $n$. The first mass 
is singled out from the rest of the KK tower as it has an extra exponential 
suppression that depends on the mass of the bulk fermion. In contrast, the 
rest of the KK tower has only a very small dependence on the mass of the 
bulk fermion through the root of the Bessel function $\xi _{n}=\xi _{n}(\nu ) 
$ which turns out to be just a linear dependence on $\nu $. 
 
\subsection{$''++''$ model} 
 
We now consider a model which exhibits Bi-gravity but does not require 
negative tension branes\cite{Kogan:2001vb}. It is built using two positive 
tension branes and leads to $AdS$ in four dimensions. A discussion of this 
model appears in Appendix B.  
 
The discussion of the localisation of spin 0 fields in the $^{\prime \prime 
}++^{\prime \prime }$ case follows similar lines to that of the $^{\prime 
\prime }+-+^{\prime \prime }$ case. In the absence of any mass term for the 
scalar the potential has the form given in eq.(80) with $\nu =\frac{3}{2}$. 
Thus again the spectrum of the scalar KK tower is identical to that of the 
graviton. However the structure changes on the addition of a five 
dimensional mass term for the scalar. If one adds a constant bulk mass term 
there is no longer a zero mode and the ultralight state is also lost. In 
this case, however, it is not possible to recover the zero mode and light 
states by adding a boundary term corresponding to coupling to boundary 
sources. The reason is that a boundary term is no longer equivalent to a 
term proportional to $\sigma ^{\prime \prime }$ (\textit{c.f.}~Appendix B). As a 
result, up to the constant bulk term we have added, the bulk potential still 
has the form of eq.(80) with $\nu =\frac{3}{2}$ and, due to the constant 
bulk mass term, there is no zero mode. We see that the multi-localisation by 
a constant bulk plus brane mass term was special to the case with negative 
tension branes. If one is to achieve the same in the case without negative 
tension branes it is necessary to add a mass term of the form given in 
eq.(6) which cannot now be interpreted as a five dimensional bulk mass term 
plus a coupling of the scalar field to boundary sources. Note that if one  
\textit{\ does} choose a scalar mass term of the form given in eq.(6) the 
remainder of the discussion applies to the $^{\prime \prime }++^{\prime 
\prime }$ case too and one can generate the multilocalised scalar field 
configurations discussed above. 
 
Of course the question is whether such a scalar mass term can be justified. 
As we will discuss in Section 9 supergravity can generate such mass terms in 
some, but not all, cases. For the remainder it seems unlikely as $\sigma 
^{\prime }$ and $\sigma ^{\prime \prime }$ are related to the metric and the 
underlying geometry of the compactification and it is difficult to see why a 
mass term of the form given in eq.(6) should arise. One possible explanation 
may follow if one can realise the ideas of reference \cite{Kehagias:2000au}. 
In this case the geometry of compactification is driven by scalar field 
vacua with kink profiles along the extra dimension. Perhaps the coupling to these scalar fields will induce a mass term of 
the form given in eq.(6). 
 
 
\section{Multi-Localization of spin $\frac{1}{2}$ field} 
 
As has been shown in Ref.\cite{Mouslopoulos:2001uc} multi-localization can 
appear also to spin $\frac{1}{2}$ fields with appropriate mass terms. Here 
for completeness we briefly review this case. The $AdS_{5}$ background 
geometry localizes the chiral zero mode on negative tension branes. However 
the addition of a mass term \cite{Rubakov:1983bb,Jackiw:1976fn} can alter 
the localization properties of the fermion so that it is localized on 
positive tension branes. The starting point again will be the action for a 
spin $\frac{1}{2}$ particle in the curved five dimensional background of 
eq.(2):  
\begin{equation} 
S=\int d^4 x \int dy \sqrt{G} \{ E^{A}_{\alpha}\left[ \frac{i}{2} \bar{\Psi} 
\gamma^{\alpha} \left( \overrightarrow{{\partial}_{A}}-\overleftarrow{%
\partial_{A}} \right) \Psi + \frac{\omega_{bcA}}{8} \bar{\Psi} \{ 
\gamma^{\alpha},\sigma^{bc} \} \Psi \right] - m(y) \bar{\Psi}\Psi \} 
\end{equation} 
where $G=det(G_{AB})=e^{-8\sigma(y)}$. Given the convention of of eq.(2) we 
adopt the ``mostly hermitian'' representation of the Dirac matrices. The 
four dimensional representation of the Dirac matrices is chosen to be $%
\gamma^{a}=( \gamma^{\mu},~~\gamma^{5} )$ with $(\gamma^{0})^{2}=-1,~ 
(\gamma^{i})^{2}=1,~ (\gamma^{5})^{2}=1$. We define $\Gamma^{M}=E^{M}_{a}%
\gamma^{a}$ and thus we have $\{\gamma^{a},\gamma^{b}\}=2 \eta^{ab}$ and $%
\{\Gamma^{a},\Gamma^{b}\}=2 g^{ab}(y)$, where $\eta^{ab}=diag(-1,1,1,1,1)$. 
The vielbein is given by  
\begin{equation} 
E^{A}_{\alpha}=diag(e^{\sigma(y)},e^{\sigma(y)},e^{\sigma(y)},e^{%
\sigma(y)},1) 
\end{equation} 
 
As in Ref.\cite{Mouslopoulos:2001uc}, we 
choose the mass term to have a (multi-) kink profile $m(y)=\frac{\sigma 
^{\prime}(y)}{k}$. It is convenient to write the action in terms of the 
fields: $\Psi_{R}$ and $\Psi_{L}$ where $\Psi_{R,L}=\frac{1}{2}(1\pm 
\gamma_{5})\Psi$ and $\Psi=\Psi_{R}+\Psi_{L}$. The action becomes:  
\begin{eqnarray} 
S=\int d^4 x \int dy \{ e^{-3\sigma}\left( \bar{\Psi}_{L}i \gamma^{\mu} 
\partial_{\mu} \Psi_{L} + \bar{\Psi}_{R}i \gamma^{\mu} \partial_{\mu} 
\Psi_{R} \right) - e^{-4\sigma} m \left(\frac{\sigma^{\prime}(y)}{k}\right) 
\left( \bar{\Psi}_{L}\Psi_{R} + \bar{\Psi}_{R}\Psi_{L} \right)  \nonumber \\ 
-\frac{1}{2}\left[ \bar{\Psi}_{L} 
(e^{-4\sigma}\partial_{y}+\partial_{y}e^{-4\sigma} ) \Psi_{R} - \bar{\Psi}%
_{R}(e^{-4\sigma}\partial_{y}+\partial_{y}e^{-4\sigma} ) \Psi_{L} \right] 
\end{eqnarray} 
writing $\Psi_{R}$ and $\Psi_{L}$ in the form:  
\begin{equation} 
\Psi_{R,L}(x,y)=\sum_{n}\psi^{R,L}_{n}(x)e^{2\sigma(y)}f_{n}^{R,L}(y) 
\end{equation} 
the action can be brought in the form  
\begin{equation} 
S=\sum_{n} \int d^4 x \{\bar{\psi}_{n}(x) i \gamma^{\mu} \partial_{\mu} 
\psi_{n}(x) - m_{n}\bar{\psi}_{n}(x) \psi_{n}(x) \} 
\end{equation} 
provided the wavefunctions obey the following equations  
\begin{eqnarray} 
\left( -\partial_{y} +m\frac{\sigma^{\prime}(y)}{k}%
\right)f^{L}_{n}(y)=m_{n}e^{\sigma(y)}f^{R}_{n}(y)  \nonumber \\ 
\left( \partial_{y} +m\frac{\sigma^{\prime}(y)}{k}%
\right)f^{R}_{n}(y)=m_{n}e^{\sigma(y)}f^{L}_{n}(y) 
\end{eqnarray} 
and the orthogonality relations (taking account of the $Z_{2}$ symmetry):  
\begin{equation} 
\int_{-L}^{L} dy e^{\sigma(y)} {f^{L}}^{*}_{m}(y) f^{L}_{n}(y)=\int_{-L}^{L} 
dy e^{\sigma(y)} {f^{R}}^{*}_{m}(y) f^{R}_{n}(y)=\delta_{mn} 
\end{equation} 
where we assume that the length of the orbifold is $2L$. 
 
We solve the above system of coupled differential equations by substituting $%
f^{L}_{n}(y)$ from the second in the first equation. Thus we end up with a 
second order differential equation, which can always be brought to a 
Schr\"ondiger form by a convenient coordinate transformation from y to z 
coordinates related through $\frac{dz}{dy}=e^{\sigma(y)}$. This gives the 
differential equation of the form:  
\begin{equation} 
\left\{- \frac{1}{2}\partial_z^2+V_{R}(z)\right\}\hat{f}^{R}_{n}(z)=\frac{%
m_n^2}{2}\hat{f}^{R} _{n}(z) 
\end{equation} 
with potential  
\begin{equation} 
\hspace*{0.5cm} V_{R}(z)=\frac{\nu(\nu+1)(\sigma^{\prime}(y))^{2}}{2[g(z)]^2}%
- \frac{\nu}{2[g(z)]^2}\sigma^{\prime\prime}(y) 
\end{equation} 
Here $\hat{f}^{R}_{n}(z)=f^{R}_{n}(y)$ and we have defined $\nu\equiv\frac{m%
}{k}$ and $g(z)\equiv e^{\sigma(y)}$. The left handed wavefunctions are 
given by \footnote{%
Note that it can be shown that the left-handed component obeys also a 
similar Schr\"odinger equation with $V_{L}(z)=\frac{\nu(\nu-1)k^{2}}{%
2[g(z)]^2}+ \frac{\nu}{2[g(z)]^2}\sigma^{\prime\prime}(y)$ which is given by  
$V_{R}$ with $\nu \rightarrow -\nu$.}:  
\begin{equation} 
f^{L}_{n}(y)= \frac{e^{-\sigma(y)}}{m_{n}} \left( \partial_{y} +m\frac{%
\sigma^{\prime}(y)}{k}\right)f^{R}_{n}(y) 
\end{equation} 
For $\nu=\frac{3}{2}$ that the form of eq.(26) is  exactly the same as that 
satisfied by the graviton. The solution has the form ${\hat{f}}%
_{n}^{R}(z)\propto e^{-\nu \sigma (y)}. $ From equation (24) we see that $%
f^{R}(y)e^{\sigma (y)/2}\propto e^{(1/2-\nu )\sigma (y)}$ is the 
appropriately normalised wavefunction. 
 
There are three regions of localization: For $\nu<\frac{1}{2}$ the zero mode 
is localized on negative tension branes, for $\nu=\frac{1}{2}$ it is not 
localized and for $\nu>\frac{1}{2}$ it is localized on positive tension 
branes. The study of the spectrum of the above differential equation 
provides the spectrum (for $\nu>\frac{1}{2}$): For the first KK state we 
find (for the symmetric configuration)  
\begin{equation} 
m_1=\sqrt{4 {\nu}^2 -1 } ~kw~ e^{-(\nu+\frac{1}{2}) x} 
\end{equation} 
and for the rest of the tower  
\begin{equation} 
m_{n+1}= \xi_n ~ kw~ e^{-x} ~~~~~~n=1,2,3, \ldots 
\end{equation} 
where $\xi_{2i+1}$ is the $(i+1)$-th root of $J_{\nu-\frac{1}{2}}(x)$ ($%
i=0,1,2, \ldots$) and $\xi_{2i}$ is the $i$-th root of $J_{\nu+\frac{1}{2}%
}(x)$ ($i=1,2,3, \ldots$). The above approximations become better away from 
the $\nu=\frac{1}{2}$, $x=0$ and for higher KK levels $n$. The first mass is 
manifestly singled out from the rest of the KK tower as it has an extra 
exponential suppression that depends on the mass of the bulk fermion. By 
contrast the rest of the KK tower has only a very small dependence on the 
mass of the bulk fermion thought the root of the Bessel function $%
\xi_{n}=\xi_{n}(\nu)$ which turns out to be just a linear dependence in $\nu$%
. The special nature of the first KK state appears not only in the 
characteristics of the mass spectrum but also in its coupling behaviour. As 
it was shown in Ref.\cite{Mouslopoulos:2001uc} the coupling to matter of the 
right-handed component of the first KK state is approximately constant 
(independent of the separation of the positive branes) 
 
It is instructive to examine the localization behaviour of the modes as the 
separation between the two $\delta $-function potential wells increases. In 
the following we assume that $\nu >\frac{1}{2}$ so that multi-localization 
is realized. In the case of infinite separation we know that each potential 
well supports a single chiral massless zero mode and that the rest of the 
massive modes come with Dirac mass terms. This raises an interesting 
question: How from the original configuration with finite size which has 
only one chiral mode do we end up with a configuration that has two chiral 
modes ? The answer to this question is found by examining the localization 
properties of the first special KK mode. From Figs (3-5) we see that the 
right-handed component of the first KK state (dashed line) is localized on 
the positive tension brane whereas the left-handed component (dotted line) 
is localized in the central region of the configuration. As we increase the 
distance separating the two potential wells the right-handed component 
remains localized on the positive tension branes whereas the left-handed 
state starts to spread along the extra dimension. We see that the second 
chiral mode that appears in the infinite separation limit is the 
right-handed component of the first KK state. This is possible since in that 
limit the left-handed component decouples (since it spreads along the 
infinite extra dimension). In this limit, the chiral zero mode that each 
potential well supports can be considered as the combination of the zero 
mode of the initial configuration and the massless limit of the first KK 
state. 
 
Our discussion of fermion localisation applies also to models of the ``$++"$ 
type without negative tension branes. The major difference is that the 
fermion mass term are no-longer constant in the bulk. As in the case of the 
scalars it remains to be seen whether such mass terms actually arise in 
models in which the geometry is determined by non-trivial scalar field 
configurations. 
 
 
\section{Localization and Multi-Localization of spin $1$ field} 
 
We now turn to the study of an Abelian gauge field. In the context of string 
theory it is natural to have gauge fields living in their world-volume of 
D-branes (these gauge fields emerge from open strings ending on the 
D-branes). However, in the case of a domain wall it turns out that it is 
difficult to localize gauge bosons in a satisfactory way. The problem has 
been addressed by several authors Ref.\cite{Dvali:1997xe,Dvali:2001rx,Kehagias:2000au,Shaposhnikov:2001nz,Tachibana:2001xq}. In this section we argue that the localization of gauge boson fields is 
indeed technically possible for particular forms of its five dimensional 
mass term (a similar mass term has been recently considered by \cite{Ghoroku:2001zu}). Our starting point is the Lagrangian for an Abelian gauge boson 
in five dimensions:  
\begin{equation} 
S=\int d^{4}x\int dy\sqrt{G}\left[ -\frac{1}{4}G^{MK}~G^{NL}~F_{MN}~F_{KL}-%
\frac{1}{2}\alpha (\sigma ^{\prime }(y))^{2}A_{M}A^{M}-\frac{1}{2}\beta 
\sigma ^{\prime \prime }(y)A_{\mu }A^{\mu }\right]  
\end{equation} 
where $F_{MN}=\partial _{M}A_{N}-\partial _{N}A_{M}$. Again we have assumed 
a mass term allowed by the symmetries of the action of the form:  
\begin{equation} 
m^{2}=\alpha (\sigma ^{\prime }(y))^{2}+\beta \sigma ^{\prime \prime }(y) 
\end{equation} 
Of course it is important to be able to generate the above mass term in a 
gauge invariant way. This can be readily done through the inclusion in the 
Lagrangian of the term:  
\begin{equation} 
\left( \alpha (\sigma ^{\prime }(y))^{2}+\beta \sigma ^{\prime \prime 
}(y)\right) \left((D^{M}\phi )^{\ast }(D_{M}\phi )-V(\phi )\right) 
\end{equation} 
Here we have added a five dimensional charged Higgs field, $\phi $. If the 
potential $V(\phi )$ triggers a vacuum expectation value for $\phi $, it 
will spontaneously break gauge invariance both in the bulk and on the brane 
and generate a vector mass term of the required form. The resulting action 
(in the gauge $A_{5}=0$)  is  
\begin{equation} 
S=\int d^{4}x\int dy\sqrt{\hat{G}}\left[ -\frac{1}{4}\hat{G}^{\mu \kappa }~%
\hat{G}^{\nu \lambda }~F_{\mu \nu }~F_{\kappa \lambda }-\frac{1}{2}%
e^{-2\sigma (y)}(\partial _{5}A_{\nu })(\partial _{5}A_{\lambda })\hat{G}%
^{\nu \lambda }-\frac{1}{2}m^{2}A_{\mu }A^{\mu }\right]  
\end{equation} 
where:  
\begin{equation} 
m^{2}=\alpha (\sigma ^{\prime }(y))^{2}+\beta \sigma ^{\prime \prime }(y) 
\end{equation} 
Performing the KK decomposition  
\begin{equation} 
A^{\mu }(x,y)=\sum_{n}A_{n}^{\mu }(x)f_{n}(y) 
\end{equation} 
this can be brought in the familiar action form for massive spin 1 particles 
propagating in flat space-time  
\begin{equation} 
S=\sum_{n}\int d^{4}x\left[ -\frac{1}{4}\eta _{\mu \kappa }~\eta _{\nu 
\lambda }~F_{n}^{\mu \nu }~F_{n}^{\kappa \lambda }-\frac{1}{2}%
m_{n}^{2}A_{n}^{\mu }A_{n}^{\nu }\right]  
\end{equation} 
provided that $f_{n}(y)$ satisfies the following second order differential 
equation  
\begin{equation} 
-\frac{d}{dy}\left( e^{-2\sigma (y)}\frac{df_{n}(y)}{dy}\right) 
+m^{2}e^{-2\sigma (y)}f_{n}(y)=m_{n}^{2}f_{n}(y) 
\end{equation} 
with the following orthogonality relations (taking in account the $Z_{2}$ 
symmetry):  
\begin{equation} 
\int_{-L}^{L}dy{f}_{m}^{\ast }(y)f_{n}(y)=\delta _{mn} 
\end{equation} 
where we assume that the length of the orbifold is $2L$. As before this can 
be brought to a Schr\"{o}dinger form by a redefinition of the wavefunction 
and by a convenient coordinate transformation from y to z coordinates 
related through: $\frac{dz}{dy}=e^{\sigma (y)}$. Thus we end up with the 
differential equation of the form:  
\begin{equation} 
\left\{ -\frac{1}{2}\partial _{z}^{2}+V(z)\right\} \hat{f}_{n}(z)=\frac{%
m_{n}^{2}}{2}\hat{f}_{n}(z) 
\end{equation} 
where  
\begin{equation} 
\hspace*{0.5cm}V(z)=\frac{\frac{3}{4}(\sigma ^{\prime }(y))^{2}+m^{2}}{%
2[g(z)]^{2}}-\frac{\frac{1}{2}}{2[g(z)]^{2}}\sigma ^{\prime \prime }(y) 
\end{equation} 
where  
\begin{equation} 
\hat{f}_{n}(z)=e^{-\frac{1}{2}\sigma (y)}f_{n}(y) 
\end{equation} 
and we have defined for convenience $g(z)\equiv e^{\sigma (y)}$. Let us now 
examine the localization properties of the gauge boson modes. For $m=0$ 
there exists a zero mode with wavefunction:  
\begin{equation} 
\hat{f}_{n}(z)=Ce^{-\frac{1}{2}\sigma (y)}=\frac{C}{\sqrt{g(z)}} 
\end{equation} 
where C is a normalization constant. From eq.(38) it is clear that in this 
case the appropriately normalized wavefunction $f(y)$ is constant along the 
extra dimension and thus that the gauge boson is delocalized. For $m\neq 0$ 
with $\alpha \neq 0$ and $\beta =0$ the zero mode becomes massive. We can 
recover the zero mode by allowing for the possibility of $\beta \neq 0$. In 
this case the potential can be written as  
\begin{equation} 
\hspace*{0.5cm}V(z)=\frac{\left( \alpha +\frac{3}{4}\right) (\sigma ^{\prime 
}(y))^{2}}{2[g(z)]^{2}}-\frac{\left( \frac{1}{2}-\beta \right) }{2[g(z)]^{2}}%
\sigma ^{\prime \prime }(y) 
\end{equation} 
This is of the general form given in Appendix A. A massless mode exists if $%
\alpha =\beta ^{2}-2\beta $ in which case the wavefunction ${\hat{f}}%
(z)\propto e^{(\beta -1/2)\sigma (y)}$. From equation (41) we see that $%
f(y)e^{-\sigma (y)}\propto e^{\beta \sigma (y)}$ is the appropriately 
normalised wavefunction in the interval $[-L,L]$. This is localised on the 
positive tension brane for $\beta >0$ and on the negative tension brane for $%
\beta <0$. When the condition for the zero mode is satisfied we find the 
mass of the first ultralight KK state to be given by (for the symmetric 
configuration):  
\begin{equation} 
m_{1}=\sqrt{4{\nu }^{2}-1}~kw~e^{-(\nu +\frac{1}{2})x} 
\end{equation} 
where $\nu =\frac{1}{2}-\beta $ and for the rest of the tower  
\begin{equation} 
m_{n+1}=\xi _{n}~kw~e^{-x}~~~~~~n=1,2,3,\ldots  
\end{equation} 
where $\xi _{2i+1}$ is the $(i+1)$-th root of $J_{\nu -\frac{1}{2}}(x)$ ($%
i=0,1,2,\ldots $) and $\xi _{2i}$ is the $i$-th root of $J_{\nu +\frac{1}{2}%
}(x)$ ($i=1,2,3,\ldots $). Again, the first KK is singled out from the rest 
of the KK tower as it has an extra exponential suppression that depends on 
the mass parameter $\nu $. In contrast the rest of the KK tower has only a 
very small dependence on the $\nu $ parameter thought the root of the Bessel 
function $\xi _{n}=\xi _{n}(\nu )$ which turns out to be just a linear 
dependence in $\nu $. 
 
Once again our discussion applies unchanged to the case without negative 
tension branes. Once again the difference is that the mass no longer 
corresponds to a combination of brane and constant bulk terms. Perhaps the 
origin of such terms will be better motivated in the case that the geometry 
is driven by a non-trivial vacuum configuration of a scalar field with a 
profile in the bulk such that coupling of the gauge field to it generates 
the required mass term. At present we have no indication that this should be 
the case. 
 
 
\section{Multi-Localization of spin $\frac{3}{2}$ field} 
 
In this section we consider the (multi-) localization of a spin $\frac{3}{2}$ 
particle. The starting point will be the Lagrangian for a $\frac{3}{2}$ 
particle propagating in curved background is:  
\begin{equation} 
S = - \int d^4 x \int dy \sqrt{G} \bar{\Psi}_{M} \Gamma^{MNP} \left( D_{N} +  
\frac{m}{2} \Gamma_{N} \right) \Psi_{P} 
\end{equation} 
where the covariant derivative is  
\begin{equation} 
D_{M} \Psi_{N} = \partial_{M} \Psi_{N} - \Gamma^{P}_{MN} \Psi_{P} + \frac{1}{%
2} \omega^{AB}_{M} \gamma_{AB} 
\end{equation} 
with $\gamma_{AB} = \frac{1}{4} [ \gamma_{A},\gamma_{B} ]$ and $%
\Gamma^{MNP}=\Gamma^{[M} \Gamma^{N} \Gamma^{P]}$.  The connection is given 
by  
\begin{eqnarray} 
\omega^{AB}_{M}=\frac{1}{2}~g^{PN}~{e^{[A}}_{P} \partial_{[M} {e^{B]}}_{N]} 
+ \frac{1}{4}~g^{PN}~g^{T\Sigma}~{e^{[A}}_{P} {e^{B]}}_{T} 
\partial_{[\Sigma} ~ {e^{\Gamma}}_{N]}~ e^{\Delta}_{M} ~\eta_{\Gamma \Delta} 
\end{eqnarray} 
where $\Gamma^{M}={e^{M}}_{n}\gamma^{n}$ with ${e^{M}}_{n}=diag(e^{%
\sigma(y)},e^{\sigma(y)},e^{\sigma(y)},e^{\sigma(y)},1)$. As in the case of 
the Abelian gauge boson we will assume that we generate the mass term for 
this field in a gauge invariant way. Exploiting the gauge invariance we can 
fix the gauge setting $\Psi_{5}=0$, something that simplifies considerably 
the calculations. In this case, the above action becomes  
\begin{equation} 
S = - \int d^4 x \int dy \sqrt{G} \bar{\Psi}_{\mu} \Gamma^{\mu \nu \rho} 
\left( D_{\nu} + \frac{m}{2} \Gamma_{\nu} \right) \Psi_{\rho} - \sqrt{G}  
\bar{\Psi}_{\mu} \Gamma^{\mu 5 \rho} \left( D_{5} + \frac{m}{2} \Gamma_{5} 
\right) \Psi_{\rho} 
\end{equation} 
We can simplify the above further taking in account the following 
identities:  
\begin{eqnarray} 
\Gamma^{\mu \nu \rho}&=&e^{3\sigma(y)}\gamma^{\mu \nu \rho}  \nonumber \\ 
\Gamma^{\mu 5 \rho}&=&e^{2\sigma(y)}\gamma^{\mu 5 \rho}  \nonumber \\ 
\gamma^{\mu \nu \rho} \gamma_{\nu}&=&-2\gamma^{\mu \rho}  \nonumber \\ 
\gamma^{\mu 5 \rho}&=&-\gamma^{5} \gamma^{\mu \rho}  \nonumber \\ 
\gamma^{\mu \rho} \gamma_{\mu}&=&-2 \gamma^{\rho} 
\end{eqnarray} 
Using the above we find the following simple forms for the covariant 
derivatives  
\begin{eqnarray} 
D_{\nu}&=&\partial_{\nu}-\frac{1}{2} \sigma^{\prime}(y)e^{-\sigma(y)} 
\gamma_{\nu}\gamma^{5}  \nonumber \\ 
D_{5}&=&\partial_{5} 
\end{eqnarray} 
Using the previous relations we write the action in the form  
\begin{eqnarray} 
S = -\int d^4 x \int dy e^{-\sigma(y)} \bar{\Psi}_{\mu} \gamma^{\mu \nu 
\rho} \left( \partial_{\nu} -\frac{1}{2} \sigma^{\prime}(y) e^{-\sigma(y)} 
\gamma_{\nu} \gamma^{5} + \frac{m}{2} e^{-\sigma(y)} \gamma_{\nu} \right) 
\Psi_{\rho}  \nonumber \\ 
- e^{-2\sigma(y)} \bar{\Psi}_{\mu} \gamma^{\mu 5 \rho} \left( \partial_{5} +  
\frac{m}{2} \gamma_{5} \right) \Psi_{\rho} 
\end{eqnarray} 
The above can be brought in the form  
\begin{equation} 
S= - \int d^4 x \int dy e^{-\sigma(y)} \bar{\Psi}_{\mu} \gamma^{\mu \nu 
\rho} \partial_{\nu}\Psi_{\rho} + e^{-2\sigma(y)} \bar{\Psi}_{\mu} 
\gamma^{\mu \rho} \left[ \frac{3m}{2} + \gamma ^{5} \left( \partial_{5} - 
\sigma^{\prime}(y) \right) \right] \Psi_{\rho} 
\end{equation} 
At this stage it turns out, like in the spin $\frac{1}{2}$ case, that it is 
convenient to write $\Psi_{\mu}$ in terms of $\Psi^{R}_{\mu}$ and $%
\Psi^{L}_{\mu}$ ($\Psi_{\mu}=\Psi^{R}_{\mu}+\Psi^{L}_{\mu}$) which have 
different KK decomposition:  
\begin{equation} 
\Psi_{\mu}^{R,L}(x,y)=\sum_{n}\psi^{R,L}_{\mu ~ 
n}(x)e^{\sigma(y)}f_{n}^{R,L}(y) 
\end{equation} 
Substituting the above decompositions in the action we get  
\begin{eqnarray} 
S = - \int d^4 x \int dy e^{\sigma(y)} ( \bar{\Psi}^{R}_{\mu} \gamma^{\mu 
\nu \rho} \partial_{\nu}\Psi^{R}_{\rho}+ \bar{\Psi}^{L}_{\mu} \gamma^{\mu 
\nu \rho} \partial_{\nu}\Psi^{L}_{\rho})  \nonumber \\ 
+ \bar{\Psi}^{R}_{\mu} \gamma^{\mu \rho} \left[ \frac{3m}{2} + \gamma ^{5} 
\partial_{5} \right] \Psi^{L}_{\rho}+ \bar{\Psi}^{L}_{\mu} \gamma^{\mu \rho} %
\left[ \frac{3m}{2} + \gamma ^{5} \partial_{5} \right] \Psi^{R}_{\rho} 
\end{eqnarray} 
this can be brought to the familiar action form for massive spin $\frac{3}{2} 
$ particle in flat background  
\begin{equation} 
S = \int d^4 x \{ - \bar{\Psi}_{\mu} \gamma^{\mu \nu \rho} 
\partial_{\nu}\Psi_{\rho} + m_{n} \bar{\Psi}_{\mu} \gamma^{\mu \rho} 
\Psi_{\rho} \} 
\end{equation} 
provided that $f^{R}_{n}$ and $\Psi^{L}_{n}$ satisfy that following coupled 
differential equations  
\begin{eqnarray} 
\left( -\partial_{y} +\frac{3m}{2} \frac{\sigma^{\prime}(y)}{k}%
\right)f^{L}_{n}(y)=m_{n}e^{\sigma(y)}f^{R}_{n}(y)  \nonumber \\ 
\left( \partial_{y} +\frac{3m}{2} \frac{\sigma^{\prime}(y)}{k}%
\right)f^{R}_{n}(y)=m_{n}e^{\sigma(y)}f^{L}_{n}(y) 
\end{eqnarray} 
supplied with the orthogonality relations (taking account of the $Z_{2}$ 
symmetry):  
\begin{equation} 
\int_{-L}^{L} dy e^{\sigma(y)} {f^{L}}^{*}_{m}(y) f^{L}_{n}(y)=\int_{-L}^{L} 
dy e^{\sigma(y)} {f^{R}}^{*}_{m}(y) f^{R}_{n}(y)=\delta_{mn} 
\end{equation} 
Note that the form of the above system of differential equations is 
identical to the one of spin $\frac{1}{2}$ particle provided we substitute $%
m \rightarrow \frac{3 m}{2}$. Accordingly the corresponding Schr\"odinger 
equation and thus the mass spectrum in this case is going to be the same as 
the spin $\frac{1}{2}$ case up to the previous rescaling of the mass 
parameter. 
 
 
\section{Multi-Localization of the graviton field} 
 
In this section, for completeness, we review the multi-localization scenario 
for the graviton. The gravitational field has the characteristic that it 
creates itself the background geometry in which it and the rest of the 
fields propagate. Thus, one has first to find the appropriate vacuum 
solution and then consider perturbations around this solution. In order to 
exhibit how the multi-localization appears in this case, we will again work 
with the $^{\prime\prime}+-+^{\prime\prime}$ configuration. The starting 
point is the Lagrangian  
\begin{equation} 
S=\int d^4 x \int_{-L_2}^{L_2} dy \sqrt{-G} \{-\Lambda + 2 M^3 
R\}-\sum_{i}\int_{y=L_i}d^4xV_i\sqrt{-\hat{G}^{(i)}} 
\end{equation} 
The Einstein equations that arise from this action are:  
\begin{equation} 
R_{MN}-\frac{1}{2}G_{MN}R=-\frac{1}{4M^3} \left(\Lambda G_{MN}+ \sum_{i}V_i%
\frac{\sqrt{-\hat{G}^{(i)}}}{\sqrt{-G}} \hat{G}^{(i)}_{\mu\nu}\delta_M^{\mu}%
\delta_N^{\nu}\delta(y-L_i)\right) 
\end{equation} 
using the metric ansatz of eq.(2) we find that the above equations imply 
that the function $\sigma(y)$ satisfies:  
\begin{eqnarray} 
\left(\sigma ^{\prime}\right)^2&=&k^2 \\ 
\sigma ^{\prime\prime}&=&\sum_{i}\frac{V_i}{12M^3}\delta(y-L_i) 
\end{eqnarray} 
where $k=\sqrt{\frac{-\Lambda}{24M^3}}$ is a measure of the curvature of the 
bulk. The exact form of $\sigma(y)$ depends on the brane configuration that 
we consider. For example, in the case of $^{\prime\prime}+-+^{\prime\prime}$ 
model we have  
\begin{equation} 
\sigma(y)=k\left\{L_1-\left||y|-L_1\right|\right\} 
\end{equation} 
where $L_{1}$ is the position of the intermediate brane. with the 
requirement that the brane tensions are tuned to $V_0=-\Lambda/k>0$, $%
V_1=\Lambda/k<0$, \mbox{$V_2=-\Lambda/k>0$}. In order to examine the 
localization properties of the graviton, the next step is to consider 
fluctuations around the vacuum of eq.(1). Thus, we expand the field $%
h_{\mu\nu}(x,y)$ in graviton and KK states plane waves:  
\begin{equation} 
h_{\mu\nu}(x,y)=\sum_{n=0}^{\infty}h_{\mu\nu}^{(n)}(x)f_{n}(y) 
\end{equation} 
where $\left(\partial_\kappa\partial^\kappa-m_n^2\right)h_{\mu\nu}^{(n)}=0$ 
and fix the gauge as $\partial^{\alpha}h_{\alpha\beta}^{(n)}=h_{\phantom{-}%
\alpha}^{(n) \alpha}=0$ \footnote{Note that we have ignored the presence of dilaton/radion fields associated 
with the size of the extra dimension or the positions of the branes. For 
more details see Ref.\cite{Charmousis:2000rg,Papazoglou:2001ed,Pilo:2000et,Kogan:2001qx}.}. The function $f_{n}(y) 
$ will obey a second order differential equation which after a change of 
variables ($\frac{dz}{dy}=e^{\sigma(y)}$) reduces to an ordinary 
Schr\"{o}dinger equation:  
\begin{equation} 
\left\{- \frac{1}{2}\partial_z^2+V(z)\right\}\hat{f}_{n}(z)=\frac{m_{n}^{2}}{%
2}\hat{f }^{n}(z) 
\end{equation} 
with potential  
\begin{equation} 
\hspace*{0.5cm} V(z)=\frac{\frac{15}{4}(\sigma^{\prime}(y))^2}{2[g(z)]^2}-  
\frac{\frac{3}{2}}{2[g(z)]^{2}} \sigma^{\prime\prime}(y) 
\end{equation} 
where  
\begin{equation} 
\hat{f}_{n}(z)\equiv f_{n}(y)e^{\sigma/2} 
\end{equation} 
and the function $g(z)$ as $g(z)\equiv 
k\left\{z_1-\left||z|-z_1\right|\right\}+1$, where $z_1=z(L_1)$. The study 
of the mass spectrum of reveals the following structure for the mass 
spectrum (for the symmetric configuration):  
\begin{eqnarray} 
m_1&=&2\sqrt{2}ke^{-2x} \\ 
m_{n+1}&=& \xi_n k e^{-x} ~~~~~~n=1,2,3, \ldots 
\end{eqnarray} 
where $\xi_{2i+1}$ is the $(i+1)$-th root of $J_1(x)$ ($i=0,1,2, \ldots$) 
and $\xi_{2i}$ is the $i$-th root of $J_2(x)$ ($i=1,2,3, \ldots$). Again the 
first KK state is singled out from the rest of the KK tower as its mass has 
an extra exponential suppression . 
 
Let us see now how the separability argument works in the case of the 
graviton. Starting with the familiar configuration $^{\prime\prime}+-+^{%
\prime\prime}$, the mass spectrum consists of the massless graviton, the 
ultra-light first KK state and the rest of the KK tower which are massive 
spin two particles. In the limit of infinite separation the first special KK 
mode becomes the second massless mode, according to our previous general 
discussions. However, at first sight the counting of degrees of freedom 
doesn't work: we start with a massive spin $2$ state (first KK state) which 
has five degrees of freedom and we end up with a massless mode which has 
two. It has been shown that in the case of flat spacetime the extra 
polarizations of the massive gravitons do not decouple giving rise to the 
celebrated van Dam-Veltman-Zakharov discontinuity in the propagator of a 
massive spin-2 field in the massless limit \cite{vanDam:1970vg,Zakharov}\footnote{However, in Ref.\cite{Vainshtein:1972sx,Deffayet:2001uk} was shown that in 
the presence of a source with a characteristic mass scale, there is no 
discontinuity for distances smaller that a critical one. This argument is 
also supported by the results of Refs.\cite{Higuchi:1987py,Kogan:2001uy,Porrati:2001cp} where it was shown that the 
limit is smooth in $dS_{4}$ or $AdS_{4}$ background.}. However our 
separability argument is still valid: Up to this point we have ignored the 
presence of a massless scalar mode, the radion, which is related to the 
motion of the freely moving negative tension brane. It turns out that this 
scalar field is a ghost field, that is, it enters the Lagrangian with the 
wrong kinetic term sign. It can be shown that the effect of the presence of 
this field is to exactly cancel the contribution of the extra polarizations 
of the graviton making the limit of infinite brane separation smooth. Note 
that apart from the radion there is another scalar field in the spectrum, 
the dilaton, which parameterizes the overall size of the extra dimension 
which also decouples in the above limit. 
 
As we have mentioned the problems associated with the presence of the ghost 
radion can be avoided by allowing for $AdS_{4}$ spacetime on the 3-branes. 
In this case there is no need for the negative tension brane (thus there is 
no radion field) and moreover the presence of curvature on the branes makes 
the massless limit of the massive graviton propagator smooth \cite{Kogan:2001uy,Porrati:2001cp}, meaning that 
in the the $AdS_{4}$ curved background the extra polarizations of the 
massive graviton decouple in the massless limit, in agreement with our 
separability argument. 
 
 
\section{Multi-Localization and supersymmetry} 
 
It is interesting to investigate the multi-localization in the 
supersymmetric versions of the previous models. The inclusion of 
supersymmetry is interesting in the sense that it restricts the possible 
mass terms by relating the mass parameters of fermion and boson fields. It 
is well known that $AdS$ spacetime is compatible with supersymmetry \cite{Townsend:1977qa,Deser:1977uq}. In contrast to the case of flat spacetime, 
supersymmetry in $AdS$ requires that fields belonging in the same multiplet 
have different masses. In the previous discussions on the localization of 
the fields, the mass term parameters which control the localization of the 
bulk states, are generally unconstrained. Let us now examine in more detail 
the cases of supergravity, vector supermultiplets and the hypermultiplet. 
 
\paragraph{Supergravity supermultiplet} 
 
The on-shell supergravity multiplet consists of the vielbein $e^{\alpha}_{M}$ 
, the graviphoton $B_{M}$ and the two symplectic-Majorana gravitinos $%
\Psi^{i}_{M}$ ($i=1,2$). The index $i$ labels the fundamental representation 
of the SU(2) automorphism group of the $N=1$ supersymmetry algebra in five 
dimensions. The supergravity Lagrangian in $AdS_{5}$ has the form \cite 
{Gherghetta:2000qt} (in $AdS_{5}$ background 
we can set $B_{M}=0$):  
\begin{eqnarray} 
S_{5}=-\frac{1}{2} \int d^{4}x \int dy \sqrt{-g} \Bigl[ M_{5}^{3} \Big\{ R+ 
i \bar{\Psi}^{i}_{M} \gamma^{MNR} D_{N} \Psi_{R}^{i} -i \frac{3}{2} 
\sigma^{\prime}(y) \bar{\Psi}^{i}_{M}\sigma^{MN}(\sigma_{3})^{ij} 
\Psi^{j}_{N} \Big\}  \nonumber \\ 
+2\Lambda -\frac{\Lambda}{k^2} \sigma^{\prime\prime}(y) \Bigr] 
\end{eqnarray} 
where $\gamma^{MNR}\equiv \sum_{perm} \frac{(-1)^{p}}{3!} \gamma^{M} 
\gamma^{N} \gamma^{R} $ and $\sigma^{MN}=\frac{1}{2}[\gamma^{M},\gamma^{N}]$%
. From the above expression, that is invariant under the supersymmetry 
transformations \cite{Gherghetta:2000qt}, we see that the 
symplectic-Majorana gravitino mass term $m=\frac{3}{2} \sigma^{\prime}(y)$ 
is such that its mass spectrum is identical to the mass spectrum of the 
graviton. This becomes clear by comparing eq.(26) for $\nu=\frac{3}{2}$ and 
eq.(66). The latter implies that, in the presence of supersymmetry, 
multi-localization of the graviton field implies multi-localization of the 
gravitinos and thus the mass spectrum of these fields will contain 
ultralight KK state(s). 
 
\paragraph{Vector supermultiplet} 
 
The on-shell field content of the vector supermultiplet $V=(V_{M},%
\lambda^{i},\Sigma)$ consists from the gauge field $V_{M}$, a 
symplectic-Majorana spinor $\lambda^{i}$, and the real scalar field $\Sigma$ 
in the adjoint representation.  
\begin{eqnarray} 
S_{5}=-\frac{1}{2} \int d^{4}x \int dy \sqrt{-g} \Bigl[ \frac{1}{2g_{5}^{2}}%
F_{MN}^{2}+(\partial_{M}\Sigma)^{2}+ i\bar{\lambda}^{i} \gamma^{M} D_{M} 
\lambda^{i} + m_{\Sigma}^{2} \Sigma^{2}+i m_{\lambda}\bar{\lambda}^{i} 
(\sigma_{3})^{ij} \lambda^{j}\Bigr] 
\end{eqnarray} 
The above Lagrangian is invariant under the supersymmetry transformations if 
the mass terms of the various fields are of the form (for more details see 
Ref.\cite{Gherghetta:2000qt}):  
\begin{eqnarray} 
m_{\Sigma}^{2}&=&-4(\sigma^{\prime}(y))^{2}+2\sigma^{\prime\prime}(y)  
\nonumber \\ 
m_{\lambda}&=& \frac{1}{2} \sigma^{\prime}(y) 
\end{eqnarray} 
Assuming that $V_{\mu}$ and $\lambda_{L}^{1}$ are even while $\Sigma$ and $%
\lambda^{2}_{L}$ odd then the mass spectrum of all the fields is identical. 
This can be easily seen if we note that for the spinors we have $\nu=\frac{1%
}{2}$, for the scalar $\alpha=-4$, $\beta=2$ and for the gauge boson $%
\alpha=0$, $\beta=0$. The even fields, for the above values of the mass 
parameters, they obey the eq.(83) of Appendix A with $\nu=\frac{1}{2}$ 
whereas the odd fields obey eq.(84) for the same value of the $\nu$ 
parameter. The mass spectrum of the two potentials is identical, apart from 
the zero modes, since they are SUSY-partner quantum mechanical potentials. 
The even fields have zero modes that are not localized (which is expected 
since the massless gauge field is not localized) in contrast to the odd 
fields that have no zero modes (they are projected out due to the boundary 
conditions). 
 
\paragraph{Hypermultiplet} 
 
The hypermultiplet $H=(H^{i},\Psi)$ consists of two complex scalar fields $%
H^{i}$ ($i=1,2$) and a Dirac fermion $\Psi$. In this case the action setup 
is:  
\begin{eqnarray} 
S_{5}=- \int d^{4}x \int dy \sqrt{-g} \Bigl[ |\partial_{M}H^{i}|^{2}+ i\bar{%
\Psi} \gamma^{M} D_{M} \Psi + m_{H^{i}}^{2} |H^{i}|^{2}+i m_{\Psi} \bar{\Psi} 
\Psi \Bigr] 
\end{eqnarray} 
Invariance under the supersymmetric transformations (see Ref.\cite 
{Gherghetta:2000qt}) demand that the mass term of the scalar and fermion 
fields has the form:  
\begin{eqnarray} 
m_{H^{1,2}}^{2}&=&(c^{2} \pm c - \frac{15}{4})(\sigma^{\prime}(y))^{2}+(%
\frac{3}{2}\mp c)\sigma^{\prime\prime}(y)  \nonumber \\ 
m_{\lambda}&=&c \sigma^{\prime}(y) 
\end{eqnarray} 
from the above we identify that $\alpha=c^2 \pm c - \frac{15}{4}$ and $\beta=%
\frac{3}{2} \mp c$ for the scalar fields. Note that $\alpha=\beta^{2}-4 \beta 
$ which implies the existence of zero mode for the symmetric scalar fields. 
Moreover, for the scalar fields we find that $\nu \equiv \frac{3}{2}- 
\beta=\pm c$ which implies that the wavefunctions (in z-coordinates) and the 
mass spectrum are identical to the Dirac fermion's. Note that we are 
assuming that $H^{1}$ and $\Psi_{L}$ are even, while $H^{2}$ and $\Psi_{R}$ 
are odd. As expected if supersymmetry is realized, multi-localization of 
scalar fields implies multi-localization of Dirac fermions and the opposite. 
 
In the five dimensional $AdS$ background,  the mass terms compatible with 
supersymmetry are not the ones that correspond to degenerate supermultiplet 
partners, but are such that all members of the supermultiplets have the same 
wavefunction behaviour and the same mass spectrum. However, in the four 
dimensional effective field theory description, the states lie in degenerate 
SUSY multiplets, as is expected since the 4D theory is flat. 
 
\section{Discussion and conclusions}

In this paper we studied the localization behaviour and the mass spectrum of 
bulk fields in various multi-brane models with localized gravity. We showed 
that the addition of appropriate mass terms controls the strength or/and the 
location of localization of the fields and moreover can induce localization 
to Abelian spin $1$ fields. The localization of all the above fields can 
resemble that of the graviton, at least in a region of the parameter space. 
This means that fields of all spins ($\leq 2$) can be 
localized on positive tension branes. The latter implies that in the context 
of multi-brane models emerges the possibility of multi-localization for all 
the previous fields with appropriate mass terms. We have shown, giving 
explicit examples, that when multi-localization is realized the above fields 
apart from the massless zero mode support ultra-light localized KK mode(s). 
 
In the simplest constructions with two positive branes, that we considered 
here, there is only one special KK state. However by adding more positive 
tension branes one can achieve more special light states. In the extreme 
example of a infinite sequence of positive branes instead of discrete 
spectrum of KK states we have continuum bands. In the previous case the 
special character of the zeroth band appears as the fact that it is well 
separated from the next. 
 
Summarizing, in this paper we pointed out some new characteristics of 
multi-brane scenarios in the case that multi-localization is realized. The 
new phenomenology reveals itself through special light and localized KK 
states. The idea of multi-localization and its relation to new interesting 
phenomenology is of course general and it should not be necessarily related 
to RS type models\footnote{In the case of gravity though, such a construction (or similar) with curved 
background is essential.}, although it finds a natural application in the 
context of these models . 
 
\vskip0.8cm 
 
\textbf{Acknowledgments:} We would like to thank Tony Gherghetta and Luigi 
Pilo for useful discussions. S.M.'s work is supported by the Greek State 
Scholarship Foundation (IKY) \mbox{No.  
8117781027}. A.P.'s work is supported by the Greek State Scholarship 
Foundation (IKY) \mbox{No. 8017711802}. This work is supported in part by 
PPARC rolling grant PPA/G/O/1998/00567, the EC TMR grant HRRN-CT-2000-00148 
and HPRN-CT-2000-00152.

\newpage

\parskip=0pt 
 
\appendix 
 
\vskip0.8cm \noindent \centerline{\Large \bf Appendix} \vskip0.4cm \noindent 
 
\section{Wavefunction Solutions} 
 
In this section we briefly discuss the form of the solutions of the Schr\"{o}%
edinger equation of the $^{\prime \prime }+-+^{\prime \prime }$ 
configuration - a configuration that exhibits multi-localization. As we have 
seen in the previous sections the general form of the potential (for field 
of any spin) of the corresponding quantum mechanical problem is of the type  
\footnote{In the following expression we have assumed that boundary mass term 
contribution is universal (its absolute value) for all the branes, \textit{i.e.} $\lambda $ is the weight of all $\delta $-functions. One might consider a 
more general case, where each $\delta $-function to have its own weight. 
However, generally the requirement of the existence of zero mode implies 
that their absolute values are equal.}:  
\begin{equation} 
\hspace*{0.5cm}V(z)=\frac{\kappa }{2[g(z)]^{2}}(\sigma ^{\prime }(y))^{2}-%
\frac{\lambda }{2[g(z)]^{2}}\sigma ^{\prime \prime }(y) 
\end{equation} 
where $\kappa $, $\lambda $ are constant parameters\footnote{%
The particular values of the parameters $\kappa $, $\lambda $ depend on the 
spin of the particle. Here we are interested in the general forms of the 
solutions. We also assume that we have already performed the appropriate 
redefinition of the wavefunction ($f(y)\rightarrow \hat{f}(z)$), which also 
depends on the spin of the particle.}. For the case of $^{\prime \prime 
}+-+^{\prime \prime }$ model we have   
\begin{eqnarray} 
(\sigma ^{\prime }(y))^{2} &=&k^{2}  \nonumber \\ 
\sigma ^{\prime \prime }(y) &=&2kg(z)\left[ \delta (z)+\delta 
(z-z_{2})-\delta (z-z_{1})\right]  
\end{eqnarray} 
$z_{1}$ and $z_{2}$ are the position of the second (negative) and the third 
(positive) brane respectively in the new coordinates ($z_{1}=z(L_{1})$ and $%
z_{2}=z(L_{2})$). The convenient choice of variables, which is universal for 
fields of all spins, is:  
\begin{equation} 
\renewcommand{\arraystretch}{1.5}z\equiv \left\{  
\begin{array}{cl} 
\frac{2e^{kL_{1}}-e^{2kL_{1}-ky}-1}{k} & y\in \lbrack L_{1},L_{2}] \\  
\frac{e^{ky}-1}{k} & y\in \lbrack 0,L_{1}] 
\end{array} 
\right. \  
\end{equation} 
(the new variable is chosen to satisfy $\frac{dz}{dy}=e^{\sigma (y)}$) and 
the function $g(z)$ is defined for convenience as $g(z)\equiv e^{\sigma 
(y)}\equiv k\left\{ z_{1}-\left| |z|-z_{1}\right| \right\} +1$. Note that in 
principle $\kappa $ and $\lambda $ are not related since the first gets 
contributions from the five dimensional bulk mass whereas the second from 
the boundary mass term. 
 
\paragraph{Even fields} 
 
Let us consider first the case that the field under consideration is even 
under the reflections $y \rightarrow -y$. In this case, it is easy to show 
that the zero mode exists only in the case that $\kappa=\nu(\nu+1)$ and $%
\lambda=\nu$ or $\lambda=-(\nu+1)$ (this is derived by imposing the boundary 
conditions coming from the $\delta$-function potentials on the massless 
solution, see below). The zero mode wavefunctions in this case have the form:%
\newline 
In the case that $\lambda=\nu$,  
\begin{equation} 
\hat{f}_{0}(z)=\frac{A}{[g(z)]^{\nu}} 
\end{equation} 
and in the case that $\lambda=-(\nu+1)$,  
\begin{equation} 
\hat{f}_{0}(z)=A^{\prime}~ [g(z)]^{\nu+1} 
\end{equation} 
Where $A$,$A^{\prime}$ are normalization constants. Note that the first is 
localized on positive tension branes whereas the second on negative tension 
branes. However, only the first choice gives the possibility of multi- 
localization on positive tension branes. Thus the existence of zero mode and 
light KK state requires $\kappa=\nu(\nu+1)$ and $\lambda=\nu$. Since we are 
interested in configurations that give rise to light KK states, the 
potential of interest is:  
\begin{equation} 
\hspace*{0.5cm} V(z)=\frac{\nu(\nu+1) k^2}{2[g(z)]^2}- \frac{\nu}{2g(z)} 2k %
\left[\delta(z)+\delta(z-z_2)-\delta(z-z_1)\right] 
\end{equation} 
For the KK modes ($m_{n} \ne 0$) the solution is given in terms of Bessel 
functions. For $y$ lying in the regions $\mathbf{A}\equiv\left[0,L_1\right]$ 
and $\mathbf{B}\equiv\left[L_1,L_2\right]$, we have:  
\begin{equation} 
{\hat{f}}_{n}\left\{ 
\begin{array}{cc} 
\mathbf{A} &  \\  
\mathbf{B} &  
\end{array} 
\right\}=\sqrt{\frac{g(z)}{k}}\left[\left\{ 
\begin{array}{cc} 
A_1 &  \\  
B_ 1 &  
\end{array} 
\right\}J_{\frac{1}{2}+\nu}\left(\frac{m_n}{k}g(z)\right)+\left\{ 
\begin{array}{cc} 
A_ 2 &  \\  
B_2 &  
\end{array} 
\right\}J_{-\frac{1}{2}-\nu}\left(\frac{m_n}{k}g(z)\right)\right] 
\end{equation} 
The boundary conditions that the wavefunctions must obey are:  
\begin{eqnarray} 
{{{\hat{f}}}_{n}}~^{\prime}({0}^{+})+\frac{k \nu}{g(0)}{{\hat{f}}}_{n}(0)=0  
\nonumber \\ 
{{\hat f}_{n}}({z_{1}}^{+})-{{\hat f}_{n}}({z_{1}}^{-})=0  \nonumber \\ 
{{\hat f}_{n}}~^{\prime}({z_{1}}^{+})-{{\hat f}_{n}}~^{\prime}({z_{1}}^{-})-%
\frac{2 k \nu}{g(z_{1})}{\hat f}_{n}(z_{1})=0  \nonumber \\ 
{{\hat f}_{n}}~^{\prime}({z_{2}}^{-})-\frac{k \nu}{g(z_{2})}{\hat f}%
_{n}(z_{2})=0 
\end{eqnarray} 
The above boundary conditions give a 4~x~4 linear homogeneous system for $%
A_{1}$, $B_{1}$, $A_{2}$ and $B_{2}$, which, in order to have a nontrivial 
solution should have vanishing determinant. This imposes a quantization 
condition from which we are able to extract the mass spectrum of the bulk 
field. 
 
\paragraph{Odd fields} 
 
In the case that the field is odd under the reflections $y \rightarrow -y$, 
there is no zero mode solution since it is not possible to make it's 
wavefunction to vanish at both boundaries. In the absence of zero mode, the 
previous restrictions between $\lambda$ and $\nu$ do not apply - and thus 
they are in principle independent. However one can ask if in this case, 
despite the absence of zero mode, a light state can exist. Indeed we can 
easily find that this can be realized for special choice of parameters: It 
can be shown that the potential  
\begin{equation} 
\hspace*{0.5cm} V_{1}(z)=\frac{\nu(\nu-1)}{2[g(z)]^2}(\sigma^{%
\prime}(y))^{2}+ \frac{\nu}{2[g(z)]^2}\sigma^{\prime\prime}(y) 
\end{equation} 
considering odd parity for the fields, gives the same spectrum (apart from 
the zero mode) with the familiar potential for fields of even parity:  
\begin{equation} 
\hspace*{0.5cm} V_{2}(z)=\frac{\nu(\nu+1)}{2[g(z)]^2}(\sigma^{%
\prime}(y))^{2}- \frac{\nu}{2[g(z)]^2}\sigma^{\prime\prime}(y) 
\end{equation} 
which according to the previous discussions supports an ultra-light special 
KK state. This is because the previous potentials are SUSY partners and as 
expected have the same spectrum apart from the zero mode. 
 
\section{Life without negative tension branes} 
 
It has been shown that the properties of the $^{\prime\prime}+-+^{\prime%
\prime}$ model (the bounce form of the ``warp'' factor), which contains a 
moving negative tension brane can be mimicked by the $^{\prime\prime}++^{%
\prime\prime}$ model, where the negative brane is absent provided that we 
allow for $AdS_{4}$ on the branes. Since the corresponding potential has two  
$\delta$-function wells that support bound states the multi-localization 
scenario appears also here. The previous results related to the localization 
properties of the various fields are valid also in this case. However, the 
presence of $AdS_{4}$ geometry on the branes, modifies the form of the 
potential of the corresponding Schr\"odinger equation and thus the details 
of the form of wavefunctions of the KK states. In this section we briefly 
discuss these modifications. 
 
As previously mentioned, the spacetime on the 3-branes must be $AdS_{4}$ (in 
contrast to the $^{\prime \prime }+-+^{\prime \prime }$ models where the 
spacetime is flat). Thus in this case the background geometry is described 
by:  
\begin{equation} 
ds^{2}=\frac{e^{-2\sigma (y)}}{(1-\frac{H^{2}x^{2}}{4})^{2}}\eta _{\mu \nu 
}dx^{\mu }dx^{\nu }+dy^{2} 
\end{equation} 
By following exactly the same steps as in the case of flat branes, again the 
whole problem is reduced to the solution of a second order differential 
equation for the profile of the KK states. The differential equation is such 
that after the dimensional reduction the five dimensional physics is 
described by a infinite tower of KK states that propagate in the $AdS_{4}$ 
background of the 3-brane. It is always possible to make the coordinate 
transformation from y coordinates to z coordinates related through: $\frac{dz%
}{dy}=A^{-1}(y)$, where $A(y)=e^{-\sigma (y)}$, and a redefinition of the 
wavefunction\footnote{The form of this redefinition depends on the spin of the field.} and bring 
the differential equation in the familiar Schr\"{o}dinger-like form:  
\begin{equation} 
\left\{ -\frac{1}{2}{\partial _{z}}^{2}+V(z)\right\} {\hat{f}}_{n}(z)=\frac{%
m_{n}^{2}}{2}{\hat{f}}_{n}(z) 
\end{equation} 
where $\hat{f}_{n}(z)$ is the appropriate redefinition of the wavefunction. 
 
For the $^{\prime\prime}++^{\prime\prime}$ model the form of the potential 
for fields of different spin is different. However, in the case that it 
admits a massless mode and an anomalously light mode it has the generic form 
given in eq.(80) that applied to the $^{\prime\prime}+-+^{\prime\prime}$ 
case. However the warp factor has a different form from the case with 
negative tension branes being given by  
\begin{equation} 
g(z)\equiv e^{\sigma (y)}=\frac{1}{\cosh (k(\left| z\right| -z_{0})} 
\end{equation} 
 
Note that in this case $\sigma ^{\prime }(y)$ is not constant in the bulk 
and $\sigma ^{\prime \prime }(y)$ is not confined to the branes. The 
massless modes, corresponding to the Schrodinger equation with this 
potential,are given by eqs.(75) and (76) as in the $^{\prime\prime}+-+^{%
\prime\prime}$ case. Note however that the constraint on the relative 
magnitude of the two terms in the potential, eq.(72), is now required when 
solving for the propagation in the bulk whereas in the case of a negative 
tension brane it came when solving for the boundary conditions. 
 
 
The zero mode wavefunction is given by:  
\begin{equation} 
\hat{f}_{0}(z)=\frac{C}{[\cos(\tilde{k}(z_{0}-|z|))]^{\nu}} 
\end{equation} 
where $C$ is the normalization factor. By considering cases with $m_{n}\neq0$%
, we find the wavefunctions for the KK tower :  
\begin{equation} 
\renewcommand{\arraystretch}{1.5}  
\begin{array}{c} 
{\hat{f}}_{n}(z)=\cos^{\nu+1}(\tilde{k}(|z|-z_{0}))\left[C_{1}~F(\tilde{a}%
_{n},\tilde{b}_{n},\frac{1}{2};\sin^{2}(\tilde{k}(|z|-z_{0})))~~~~~~~~\right. 
\\  
\left. ~~~~~~~~~~~~~~~~~~~+C_{2}~|\sin(\tilde{k}(|z|-z_{0}))|~F(\tilde{a}%
_{n}+\frac{1}{2},\tilde{b}_{n}+\frac{1}{2},\frac{3}{2};\sin^{2}(\tilde{k}%
(|z|-z_{0})))\right] 
\end{array} 
\end{equation} 
where  
\begin{eqnarray} 
\tilde{a}_{n}=\frac{\nu+1}{2}+\frac{1}{2}\sqrt{\left(\frac{m_{n}}{\tilde{k}}%
\right)^2+{\nu}^2} \cr \tilde{b}_{n}=\frac{\nu+1}{2}-\frac{1}{2}\sqrt{\left(%
\frac{m_{n}}{\tilde{k}}\right)^2+{\nu}^2 } 
\end{eqnarray} 
The boundary conditions are given by:  
\begin{eqnarray} 
{\hat{f}}_{n}~^{\prime}({0}^{+})+k \nu \tanh(k y_{0}){\hat{f}}_{n}(0)=0  
\nonumber \\ 
{\hat{f}}_{n}~^{\prime}({z_{L}}^{-})-k \nu \frac{\sinh(k(L-y_{0}))}{\cosh(k 
y_{0})}{\hat{f}}_{n}(z_{L})=0 
\end{eqnarray} 
the above conditions determine the mass spectrum of the KK states. By 
studying the mass spectrum of the KK states it turns out that it has a 
special first mode similar to the one of the $^{\prime\prime}+-+^{\prime%
\prime}$ model as expected. 
 


\begin{thebibliography}{999} 
\parskip=-6pt 
 
 
{\footnotesize 
} 
 
\bibitem{Akama:1982jy}  {\footnotesize K.~Akama, ``An Early Proposal Of 
'Brane World','' Lect.\ Notes Phys.\ \textbf{176} (1982) 267 
[hep-th/0001113]. 
} 
 
{\footnotesize 
} 
 
\bibitem{Rubakov:1983bb}  {\footnotesize V.~A.~Rubakov and 
M.~E.~Shaposhnikov, ``Do We Live Inside A Domain Wall?,'' Phys.\ Lett.\ B  
\textbf{125}, 136 (1983). 
} 
 
{\footnotesize 
} 
 
\bibitem{Visser:1985qm}  {\footnotesize M.~Visser, ``An Exotic Class Of 
Kaluza-Klein Models,'' Phys.\ Lett.\ B \textbf{159}, 22 (1985) 
[hep-th/9910093]. 
} 
 
{\footnotesize 
} 
 
\bibitem{Squires:1986aq}  {\footnotesize E.~J.~Squires, ``Dimensional 
Reduction Caused By A Cosmological Constant,'' Phys.\ Lett.\ B \textbf{167}, 
286 (1986). 
} 
 
{\footnotesize 
} 
 
{\footnotesize 
} 
 
\bibitem{Arkani-Hamed:1998rs}  {\footnotesize N.~Arkani-Hamed, S.~Dimopoulos 
and G.~Dvali, ``The hierarchy problem and new dimensions at a millimeter,'' 
Phys.\ Lett.\ B \textbf{429} (1998) 263 [hep-ph/9803315].  
} 
 
{\footnotesize 
} 
 
{\footnotesize 
} 
 
\bibitem{Antoniadis:1998ig}  {\footnotesize I.~Antoniadis, N.~Arkani-Hamed, 
S.~Dimopoulos and G.~Dvali, ``New dimensions at a millimeter to a Fermi and 
superstrings at a TeV,'' Phys.\ Lett.\ B \textbf{436} (1998) 257 
[hep-ph/9804398]. 
} 
 
{\footnotesize 
} 
 
{\footnotesize 
} 
 
\bibitem{Arkani-Hamed:1999nn}  {\footnotesize N.~Arkani-Hamed, S.~Dimopoulos 
and G.~Dvali, ``Phenomenology, astrophysics and cosmology of theories with 
sub-millimeter dimensions and TeV scale quantum gravity,'' Phys.\ Rev.\ D  
\textbf{59} (1999) 086004 [hep-ph/9807344]. 
} 
 
{\footnotesize 
} 
 
{\footnotesize 
} 
 
\bibitem{Gogberashvili:1998vx}  {\footnotesize M.~Gogberashvili, ``Hierarchy 
problem in the shell-universe model,'' hep-ph/9812296.  
} 
 
{\footnotesize 
} 
 
{\footnotesize 
} 
 
\bibitem{Randall:1999ee}  {\footnotesize L.~Randall and R.~Sundrum, ``A 
large mass hierarchy from a small extra dimension,'' Phys.\ Rev.\ Lett.\  
\textbf{83} (1999) 3370 [hep-ph/9905221]. 
} 
 
{\footnotesize 
} 
 
{\footnotesize 
} 
 
\bibitem{Randall:1999vf}  {\footnotesize L.~Randall and R.~Sundrum, ``An 
alternative to compactification,'' Phys.\ Rev.\ Lett.\ \textbf{83} (1999) 
4690 [hep-th/9906064]. 
} 
 
{\footnotesize 
} 
 
{\footnotesize 
} 
 
\bibitem{Dienes:1999sb}  {\footnotesize K.~R.~Dienes, E.~Dudas and 
T.~Gherghetta, ``Light neutrinos without heavy mass scales: A 
higher-dimensional seesaw mechanism,'' Nucl.\ Phys.\ B \textbf{557} (1999) 
25 [hep-ph/9811428]. 
} 
 
{\footnotesize 
} 
 
{\footnotesize 
} 
 
\bibitem{Arkani-Hamed:1998vp}  {\footnotesize N.~Arkani-Hamed, 
S.~Dimopoulos, G.~Dvali and J.~March-Russell, ``Neutrino masses from large 
extra dimensions,'' hep-ph/9811448. 
} 
 
{\footnotesize 
} 
 
{\footnotesize 
} 
 
\bibitem{Dvali:1999cn}  {\footnotesize G.~Dvali and A.~Y.~Smirnov, ``Probing 
large extra dimensions with neutrinos,'' Nucl.\ Phys.\ B \textbf{563} (1999) 
63 [hep-ph/9904211]. 
} 
 
{\footnotesize 
} 
 
{\footnotesize 
} 
 
\bibitem{Mohapatra:1999zd}  {\footnotesize R.~N.~Mohapatra, S.~Nandi and 
A.~Perez-Lorenzana, ``Neutrino masses and oscillations in models with large 
extra dimensions,'' Phys.\ Lett.\ B \textbf{466} (1999) 115 
[hep-ph/9907520]. 
} 
 
{\footnotesize 
} 
 
{\footnotesize 
} 
 
\bibitem{Grossman:2000ra}  {\footnotesize Y.~Grossman and M.~Neubert, 
``Neutrino masses and mixings in non-factorizable geometry,'' Phys.\ Lett.\ 
B \textbf{474} (2000) 361 [hep-ph/9912408]. 
} 




{\footnotesize 
} 
 
\bibitem{Barbieri:2000mg}  {\footnotesize R.~Barbieri, P.~Creminelli and 
A.~Strumia, ``Neutrino oscillations from large extra dimensions,'' Nucl.\ 
Phys.\ B \textbf{585} (2000) 28 [hep-ph/0002199].  
} 
 
{\footnotesize 
} 
 
{\footnotesize 
} 
 
\bibitem{Lukas:2000wn}  {\footnotesize A.~Lukas, P.~Ramond, A.~Romanino and 
G.~G.~Ross, ``Solar neutrino oscillation from large extra dimensions,'' 
Phys.\ Lett.\ B \textbf{495} (2000) 136 [hep-ph/0008049].  
} 
 
{\footnotesize 
} 

{\footnotesize 
} 
 
\bibitem{Cosme:2000ib}  {\footnotesize N.~Cosme, J.~M.~Frere, Y.~Gouverneur, 
F.~S.~Ling, D.~Monderen and V.~Van Elewyck, ``Neutrino suppression and extra 
dimensions: A minimal model,'' hep-ph/0010192.  
} 



 
{\footnotesize 
} 
 
\bibitem{Lukas:2000rg}  {\footnotesize A.~Lukas, P.~Ramond, A.~Romanino and 
G.~G.~Ross, ``Neutrino masses and mixing in brane-world theories,'' 
hep-ph/0011295. 
} 
 
 
 
{\footnotesize 
} 
 
\bibitem{Mouslopoulos:2001uc}  {\footnotesize S.~Mouslopoulos, ``Bulk 
fermions in multi-brane worlds,'' JHEP \textbf{0105} (2001) 038 
[hep-th/0103184]. 
} 
 
{\footnotesize 
} 
 
{\footnotesize 
} 
 
\bibitem{Arkani-Hamed:2000dc}  {\footnotesize N.~Arkani-Hamed and 
M.~Schmaltz, ``Hierarchies without symmetries from extra dimensions,'' 
Phys.\ Rev.\ D \textbf{61} (2000) 033005 [hep-ph/9903417].  
} 
 
{\footnotesize 
} 
 
{\footnotesize 
} 
 
\bibitem{Mirabelli:2000ks}  {\footnotesize E.~A.~Mirabelli and M.~Schmaltz, 
``Yukawa hierarchies from split fermions in extra dimensions,'' Phys.\ Rev.\ 
D \textbf{61} (2000) 113011 [hep-ph/9912265].  
} 
 
{\footnotesize 
} 
 
\bibitem{Dvali:2000ha}  {\footnotesize G.~Dvali and M.~Shifman, ``Families 
as neighbors in extra dimension,'' Phys.\ Lett.\ B \textbf{475}, 295 (2000) 
[hep-ph/0001072]. 
} 
 
{\footnotesize 
} 
 
{\footnotesize 
} 
 
\bibitem{delAguila:2000kb}  {\footnotesize F.~del Aguila and J.~Santiago, 
``Universality limits on bulk fermions,'' Phys.\ Lett.\ B \textbf{493} 
(2000) 175 [hep-ph/0008143]. 
} 
 
{\footnotesize 
} 
 
{\footnotesize 
} 
 
{\footnotesize 
} 
 
\bibitem{Goldberger:1999wh}  {\footnotesize W.~D.~Goldberger and M.~B.~Wise, 
``Bulk fields in the Randall-Sundrum compactification scenario,'' Phys.\ 
Rev.\ D \textbf{60} (1999) 107505 [hep-ph/9907218].  
} 
 
{\footnotesize 
} 
 
{\footnotesize 
} 
 
\bibitem{Mintchev:2001mf}  {\footnotesize M.~Mintchev and L.~Pilo, 
``Localization of quantum fields on branes,'' Nucl.\ Phys.\ B \textbf{592} 
(2001) 219 [hep-th/0007002]. 
} 
 
{\footnotesize  
} 
 
{\footnotesize 
} 
 
\bibitem{Bajc:2000mh}  {\footnotesize B.~Bajc and G.~Gabadadze, 
``Localization of matter and cosmological constant on a brane in anti de 
Sitter space,'' Phys.\ Lett.\ B \textbf{474} (2000) 282 [hep-th/9912232].  
} 
 
{\footnotesize 
} 
 
{\footnotesize 
} 
 
\bibitem{Chang:2000nh}  {\footnotesize S.~Chang, J.~Hisano, H.~Nakano, 
N.~Okada and M.~Yamaguchi, ``Bulk standard model in the Randall-Sundrum 
background,'' Phys.\ Rev.\ D \textbf{62} (2000) 084025 [hep-ph/9912498].  
} 
 
{\footnotesize 
} 
 
{\footnotesize 
} 
 
\bibitem{Gherghetta:2000qt}  {\footnotesize T.~Gherghetta and A.~Pomarol, 
``Bulk fields and supersymmetry in a slice of AdS,'' Nucl.\ Phys.\ B \textbf{%
586} (2000) 141 [hep-ph/0003129]. 
} 
 
{\footnotesize 
} 
 
{\footnotesize 
} 
 
\bibitem{Randjbar-Daemi:2000cr}  {\footnotesize S.~Randjbar-Daemi and 
M.~Shaposhnikov, ``Fermion zero-modes on brane-worlds,'' Phys.\ Lett.\ B  
\textbf{492} (2000) 361 [hep-th/0008079]. 
} 
 
{\footnotesize 
} 
 
{\footnotesize 
} 
 
\bibitem{Oda:2000kh}  {\footnotesize I.~Oda, ``Localization of various bulk 
fields on a brane,'' hep-th/0009074. 
} 
 
{\footnotesize 
} 
 
{\footnotesize 
} 
 
\bibitem{Kehagias:2000au}  {\footnotesize A.~Kehagias and K.~Tamvakis, 
``Localized gravitons, gauge bosons and chiral fermions in smooth spaces 
generated by a bounce,'' hep-th/0010112. 
} 
 
{\footnotesize 
} 
 
{\footnotesize 
} 
 
\bibitem{Oda:2000dd}  {\footnotesize I.~Oda, ``Localization of bulk fields 
on $AdS_{4}$ brane in $AdS_{5}$,'' hep-th/0012013.  
} 
 
{\footnotesize 
} 
 
{\footnotesize 
} 
 
\bibitem{Gherghetta:2000kr}  {\footnotesize T.~Gherghetta and A.~Pomarol,
Nucl.\ Phys.\ B {\bf 602}, 3 (2001)
[hep-ph/0012378].
} 
 
{\footnotesize 
} 
 
{\footnotesize 
} 
 
\bibitem{Oda:2000wa}  {\footnotesize I.~Oda, ``Localization of gravitino on 
a brane,'' hep-th/0008134. 
} 
 
{\footnotesize 
} 



{\footnotesize 
} 
 
\bibitem{Davoudiasl:2000tf}  {\footnotesize H.~Davoudiasl, J.~L.~Hewett and 
T.~G.~Rizzo, ``Bulk gauge fields in the Randall-Sundrum model,'' Phys.\ 
Lett.\ B \textbf{473} (2000) 43 [hep-ph/9911262].  
} 



 
{\footnotesize 
} 
 
\bibitem{Pomarol:2000ad}  {\footnotesize A.~Pomarol, ``Gauge bosons in a 
five-dimensional theory with localized gravity,'' Phys.\ Lett.\ B \textbf{486%
} (2000) 153 [hep-ph/9911294]. 
} 


 
{\footnotesize 
} 
 
\bibitem{Davoudiasl:2001wi}  {\footnotesize H.~Davoudiasl, J.~L.~Hewett and 
T.~G.~Rizzo, ``Experimental probes of localized gravity: On and off the 
wall,'' Phys.\ Rev.\ D \textbf{63} (2001) 075004 [hep-ph/0006041].  
} 
 


{\footnotesize  
} 

\bibitem{Duff:2001jk}{\footnotesize M.~J.~Duff, J.~T.~Liu and W.~A.~Sabra,
``Localization of supergravity on the brane,'' Nucl.\ Phys.\ B {\bf
605}, 234 (2001) [hep-th/0009212].
} 



{\footnotesize 
} 

{\footnotesize 
} 
 
\bibitem{Jackiw:1976fn}  {\footnotesize R.~Jackiw and C.~Rebbi, ``Solitons 
With Fermion Number $\frac{1}{2}$,'' Phys.\ Rev.\ D \textbf{13} (1976) 3398.  
} 




 
{\footnotesize 
} 
 
\bibitem{Kogan:2000wc}  {\footnotesize I.~I.~Kogan, S.~Mouslopoulos, 
A.~Papazoglou, G.~G.~Ross and J.~Santiago, ``A three three-brane universe: 
New phenomenology for the new millennium?,'' Nucl.\ Phys.\ B \textbf{584} 
(2000) 313 [hep-ph/9912552]. 
} 
 
{\footnotesize 
} 
 
{\footnotesize 
} 
 
\bibitem{Mouslopoulos:2000er}  {\footnotesize S.~Mouslopoulos and 
A.~Papazoglou, ``$^{\prime\prime}+ - +^{\prime\prime}$ brane model 
phenomenology,'' JHEP\textbf{0011} (2000) 018 [hep-ph/0003207].  
} 
 

 
{\footnotesize 
} 
 
\bibitem{Kogan:2001vb}  {\footnotesize I.~I.~Kogan, S.~Mouslopoulos and 
A.~Papazoglou, ``A new bigravity model with exclusively positive branes,'' 
Phys.\ Lett.\ B \textbf{501} (2001) 140 [hep-th/0011141].  
} 


{\footnotesize 
} 
 
{\footnotesize 
} 
 
\bibitem{Kogan:2001yr}  {\footnotesize I.~I.~Kogan, S.~Mouslopoulos, 
A.~Papazoglou and G.~G.~Ross,  
``Multigravity in six dimensions: Generating bounces with flat positive  tension branes,'' 
hep-th/0107086. 
} 
 
{\footnotesize 
} 
 
{\footnotesize 
} 
 
\bibitem{Gregory:2000jc}  {\footnotesize R.~Gregory, V.~A.~Rubakov and 
S.~M.~Sibiryakov, ``Opening up extra dimensions at ultra-large scales,'' 
Phys.\ Rev.\ Lett.\ \textbf{84} (2000) 5928 [hep-th/0002072].  
} 
 
{\footnotesize 
} 
 
{\footnotesize 
} 
 
\bibitem{Kogan:2000cv}  {\footnotesize I.~I.~Kogan and G.~G.~Ross, ``Brane 
universe and multigravity: Modification of gravity at large and small 
distances,'' Phys.\ Lett.\ B \textbf{485} (2000) 255 [hep-th/0003074].  
} 
 
{\footnotesize 
} 
 
{\footnotesize 
} 
 
\bibitem{Kogan:2000xc}  {\footnotesize I.~I.~Kogan, S.~Mouslopoulos, 
A.~Papazoglou and G.~G.~Ross, ``Multi-brane worlds and modification of 
gravity at large scales,'' Nucl.\ Phys.\ B \textbf{595} (2001) 225 
[hep-th/0006030]. 
} 
 
{\footnotesize 
} 

 
{\footnotesize 
} 
 
{\footnotesize 
} 
 
\bibitem{Karch:2001ct}  {\footnotesize A.~Karch and L.~Randall, ``Locally 
localized gravity,'' Int.\ J.\ Mod.\ Phys.\ A \textbf{16} (2001) 780 
[hep-th/0011156]. 
} 
 
{\footnotesize 
} 
 
{\footnotesize 
} 
 
\bibitem{Miemiec:2000eq}  {\footnotesize A.~Miemiec, ``A power law for the 
lowest eigenvalue in localized massive gravity,'' hep-th/0011160.  
} 
 
{\footnotesize 
} 
 
{\footnotesize 
} 
 
\bibitem{Schwartz:2001ip}  {\footnotesize M.~D.~Schwartz, ``The emergence of 
localized gravity,'' Phys.\ Lett.\ B \textbf{502} (2001) 223 
[hep-th/0011177]. 
} 
 
{\footnotesize 
} 
 
{\footnotesize 
} 
 
\bibitem{Karch:2001cw}  {\footnotesize A.~Karch and L.~Randall, ``Localized 
Gravity in String Theory,'' hep-th/0105108. 
} 




{\footnotesize 
}

\bibitem{Gorsky:2000rz}{\footnotesize A.~Gorsky and K.~Selivanov,
Phys.\ Lett.\ B {\bf 485}, 271 (2000)
[hep-th/0005066].
}



{\footnotesize 
} 
 
\bibitem{Dvali:1997xe}  {\footnotesize G.~Dvali and M.~Shifman, ``Domain 
walls in strongly coupled theories,'' Phys.\ Lett.\ B \textbf{396} (1997) 64 
[Erratum-ibid.\ B \textbf{407} (1997) 452] [hep-th/9612128].  
} 
 
{\footnotesize 
} 
 
\bibitem{Dvali:2001rx}  {\footnotesize G.~Dvali, G.~Gabadadze and 
M.~Shifman, ``(Quasi)localized gauge field on a brane: Dissipating cosmic 
radiation to extra dimensions?,'' Phys.\ Lett.\ B \textbf{497} (2001) 271 
[hep-th/0010071]. 
} 


{\footnotesize 
} 
 
\bibitem{Shaposhnikov:2001nz}  {\footnotesize M.~Shaposhnikov and 
P.~Tinyakov, ``Extra dimensions as an alternative to Higgs mechanism?,'' 
hep-th/0102161. 
} 

 
{\footnotesize 
} 
 
\bibitem{Tachibana:2001xq}  {\footnotesize M.~Tachibana, ``Comment on 
Kaluza-Klein Spectrum of Gauge Fields in the Bigravity Model,'' 
hep-th/0105180. 
} 
 



{\footnotesize 
} 
 
\bibitem{Ghoroku:2001zu}  {\footnotesize K.~Ghoroku and A.~Nakamura,  
``Massive vector trapping as a gauge boson on a brane,'' 
hep-th/0106145. 
} 



{\footnotesize  
} 

\bibitem{Charmousis:2000rg}{\footnotesize C.~Charmousis, R.~Gregory and V.~A.~Rubakov,
``Wave function of the radion in a brane world,''
Phys.\ Rev.\ D {\bf 62}, 067505 (2000)
[hep-th/9912160].
}


{\footnotesize 
} 
 
\bibitem{Pilo:2000et}  {\footnotesize L.~Pilo, R.~Rattazzi and A.~Zaffaroni, 
``The fate of the radion in models with metastable graviton,'' JHEP\textbf{%
0007} (2000) 056 [hep-th/0004028]. 
} 



{\footnotesize  
} 


\bibitem{Papazoglou:2001ed}{\footnotesize A.~Papazoglou,
``Dilaton tadpoles and mass in warped models,''
Phys.\ Lett.\ B {\bf 505}, 231 (2001)
[hep-th/0102015].
}


{\footnotesize 
} 
 
\bibitem{Kogan:2001qx}  {\footnotesize I.~I.~Kogan, S.~Mouslopoulos, 
A.~Papazoglou and L.~Pilo, ``Radion in Multibrane World,'' hep-th/0105255.  
} 



{\footnotesize 
}

\bibitem{vanDam:1970vg}   {\footnotesize H.~van Dam and M.~Veltman,  
Nucl.\ Phys.\ B {\bf 22}, 397 (1970).  
}
 

{\footnotesize 
}

\bibitem{Zakharov} {\footnotesize V.I. Zakharov, JETP
Lett. \textbf{12} (1970) 312.
}
 
{\footnotesize 
} 
 
\bibitem{Vainshtein:1972sx}  {\footnotesize A.~I.~Vainshtein, ``To The 
Problem Of Nonvanishing Gravitation Mass,'' Phys.\ Lett.\ B \textbf{39} 
(1972) 393. 
} 
 
{\footnotesize 
} 
 
\bibitem{Deffayet:2001uk}  {\footnotesize C.~Deffayet, G.~Dvali, 
G.~Gabadadze and A.~Vainshtein,  
``Nonperturbative continuity in graviton mass versus perturbative  discontinuity,'' 
hep-th/0106001. 
} 
 
{\footnotesize 
} 
 
\bibitem{Higuchi:1987py}  {\footnotesize A.~Higuchi,  
``Forbidden Mass Range For Spin-2 Field Theory In De Sitter Space-Time,'' 
Nucl.\ Phys.\ B \textbf{282} (1987) 397. 
} 
 
{\footnotesize 
} 
 
\bibitem{Kogan:2001uy}  {\footnotesize I.~I.~Kogan, S.~Mouslopoulos and 
A.~Papazoglou, ``The m $\to$ 0 limit for massive graviton in $dS_{4}$ and $%
AdS_{4}$: How to circumvent the van Dam-Veltman-Zakharov discontinuity,'' 
Phys.\ Lett.\ B \textbf{503} (2001) 173 [hep-th/0011138].  
} 
 
{\footnotesize 
} 
 
\bibitem{Porrati:2001cp}  {\footnotesize M.~Porrati, ``No van 
Dam-Veltman-Zakharov discontinuity in AdS space,'' Phys.\ Lett.\ B \textbf{%
498} (2001) 92 [hep-th/0011152]. 
} 
 
{\footnotesize 
} 
 
{\footnotesize 
} 
 
\bibitem{Townsend:1977qa}  {\footnotesize P.~K.~Townsend, ``Cosmological 
Constant In Supergravity,'' Phys.\ Rev.\ D \textbf{15} (1977) 2802.  
} 
 
{\footnotesize 
} 
 
{\footnotesize 
} 
 
\bibitem{Deser:1977uq}  {\footnotesize S.~Deser and B.~Zumino, ``Broken 
Supersymmetry And Supergravity,'' Phys.\ Rev.\ Lett.\ \textbf{38} (1977) 
1433. 
} 
 
 
 




 
 
 
 
 
 
 
 
 
 
 
 
 
 
 
 

\end{thebibliography}
\end{document}